\documentclass[12pt,letterpaper]{JHEP3}

\usepackage{amscd,amsmath,amssymb,amsfonts,xspace,mathrsfs}
\usepackage{epsfig}

\numberwithin{equation}{section}

\def\varpi{t}

\def\Im{\,{\rm Im}\,}
\def\Re{\,{\rm Re}\,}

\def\({\left(}
\def\){\right)}
\def\[{\left[}
\def\]{\right]}
\def\hf{{1\over 2}}

\newcommand{\de}{\mathrm{d}}

\newcommand{\I}{\mathrm{i}}

\newcommand{\p}{\partial}

\newcommand{\cV}{\mathcal{V}}

\newcommand{\cC}{\mathcal{C}}

\newcommand{\cK}{\mathcal{K}}
\newcommand{\cM}{\mathcal{M}}

\newcommand{\cN}{\mathcal{N}}
\newcommand{\cE}{\mathcal{E}}
\newcommand{\cX}{\mathcal{X}}
\newcommand{\CX}{\mathcal{X}}

\DeclareSymbolFont{AMSa}{U}{msa}{m}{n}
\DeclareSymbolFont{AMSb}{U}{msb}{m}{n}
\DeclareMathSymbol{\fieldR}{\mathalpha}{AMSb}{"52}

\newcommand{\N}{{\mathcal N}}
\newcommand{\kahler}{{K\"ahler}\xspace}
\newcommand{\hk}{{hyperk\"ahler}\xspace}
\newcommand{\qk}{{quaternion-K\"ahler}\xspace}

\newcommand{\cZ}{\mathcal{Z}}
\newcommand{\cI}{\mathcal{I}}
\newcommand{\cO}{\mathcal{O}}

\newcommand{\cU}{\mathcal{U}}

\newcommand{\pa}{\partial}
\newcommand{\nn}{\nonumber}

\newcommand{\IR}{\mathbb{R}}
\newcommand{\IC}{\mathbb{C}}
\newcommand{\IZ}{\mathbb{Z}}

\newcommand{\tzeta}{\tilde\zeta}

\newcommand{\txi}{\tilde\xi}
\newcommand{\tc}{\tilde c}

\newcommand{\CP}{\IC P^1}

\def\bea{\begin{eqnarray}}
\def\eea{\end{eqnarray}}
\def\be{\begin{equation}}
\def\ee{\end{equation}}
\def\ba{\begin{align}}
\def\ea{\end{align}}
\def\bse{\begin{subequations}}
\def\ese{\end{subequations}}

\def\bi{\bar \imath}
\def\bj{\bar \jmath}
\def\ba{\bar a}

\def\bY{\bar Y}

\def\bF{\bar F}

\def\ui#1{^{[#1]}}

\def\txii#1{{\tilde\xi}^{[#1]}}

\def\ai#1{{\alpha}^{[#1]}}

\def\xii#1{\xi_{[#1]}}
\def\rhoi#1{\rho_{[#1]}}

\def\Hij#1{H^{[#1]}}

\def\XXint#1#2#3{{\setbox0=\hbox{$#1{#2#3}{\int}$}
\vcenter{\hbox{$#2#3$}}\kern-.5\wd0}}

\def\hHij#1{H^{[#1]}}
\newcommand{\hU}{\hat{\mathcal{U}}}
\newcommand{\hCX}{\mathcal{X}}

\def\cij#1{c}
\def\ci#1{c}

\newcommand{\cY}{\mathcal{Y}}

\def\kk{{\bf k}}

\def\htt{{\mathtt t}}
\def\htp{\htt_+}
\def\htm{\htt_-}
\def\htpm{\htt_\pm}

\def\tinv{s}
\def\tinvp{\tinv_+}
\def\tinvm{\tinv_-}
\def\tinvpm{\tinv_\pm}

\def\rhoip{\rho_{c,d}^+}
\def\rhoim{\rho_{c,d}^-}
\def\rhoipm{\rho_{c,d}^\pm}

\def\rhoip{\rho_{c,d}^+}
\def\rhoim{\rho_{c,d}^-}
\def\rhoipm{\rho_{c,d}^\pm}

\def\Phip{\Phi_{c,d}^+}
\def\Phim{\Phi_{c,d}^-}
\def\Phipm{\Phi_{c,d}^\pm}

\def\Cf{\cC^{c,d}}
\def\Cfp{\Cf_+}
\def\Cfm{\Cf_-}
\def\Cfpm{\Cf_\pm}

\def\rhocd{\varrho_{c,d}}

\def\Mcl{\cM_{0}}
\def\MH{\cM_{\rm HM}}
\def\MV{\cM_{\rm VM}}
\def\Mcm{\cM_{\text{c-map}}}

\def\KK{{\mathscr M}_K}
\def\SK{{\mathscr M}_{\rm sk}}

\def\gl#1{{\rm g}_{#1}}
\def\glt#1{{\rm g}_{#1}[\varpi]}

\def\Fi{\mathscr R}

\title{Modularity, Quaternion-K\"ahler spaces and Mirror Symmetry
}

\author{Sergei Alexandrov and Sibasish Banerjee
\\
{\it Universit\'e Montpellier 2, Laboratoire Charles Coulomb UMR 5221, F-34095,
Montpellier, France}\\

\vspace*{2mm} {\tt e-mail:
\email{salexand@univ-montp2.fr},
\email{sibasishbanerjee@live.in}
}

\vspace*{-3mm}

}

\abstract{We provide an explicit twistorial construction of quaternion-K\"ahler manifolds
obtained by deformation of c-map spaces and carrying an isometric action of the modular group $SL(2,\IZ)$.
The deformation is not assumed to preserve any continuous isometry and therefore this construction presents
a general framework for describing NS5-brane instanton effects in string compactifications with $N=2$ supersymmetry.
In this context the modular invariant parametrization of twistor lines found in this work yields
the complete non-perturbative mirror map between type IIA and type IIB physical fields.}

\begin{document}

\section{Introduction}
\label{secIntr}

Modular invariance plays a prominent role both in physics and mathematics.
Especially important it appears to be in supersymmetric gauge and string theories where
the S-transformation of the modular group $SL(2,\IZ)$ generates the weak/strong coupling duality.
Furthermore, considering low energy effective actions of these theories, one encounters a beautiful interplay
between modular invariance and the geometry of special holonomy manifolds. For instance, this interplay
was the basis for the famous Seiberg-Witten solution \cite{Seiberg:1994rs}, describing the moduli space
of 4d $N=2$ gauge theories encoded in the metric on a certain rigid special K\"ahler manifold,
whereas compactifying them further on a circle, one falls into the realm of \hk geometry \cite{Seiberg:1996nz}.

In this work we study the interconnection of modular invariance with \qk (QK) geometry.
Our prime motivation comes from string theory where the situation relevant to our problem arises
in type IIB string theory compactified on a Calabi-Yau (CY) threefold. In this case the low energy effective action
is known to be described by the metric on the vector multiplet (VM) and hypermultiplet (HM) moduli spaces, $\MV$ and $\MH$,
and the latter is known to be \qk \cite{Bagger:1983tt} and to carry an isometric action of $SL(2,\IZ)$
\cite{Green:1997tv,RoblesLlana:2006is}.
Together with additional discrete symmetries and the information provided by perturbative string calculations,
these conditions are strong enough to hope that the metric on $\MH$ can be explicitly found.
Although significant progress has been achieved in this direction in recent years
(see \cite{Alexandrov:2011va,Alexandrov:2013yva} for recent reviews), the complete solution is still unknown.

Our aim here is to provide an explicit description of a certain class of QK manifolds which are restricted
to have an isometric action of $SL(2,\IZ)$, so that $\MH$ appears as a particular member of this class.
But first we should explain what this class is and what we mean by ``explicit description".

The QK manifolds we are going to consider are all deformations of the space which will be denoted by $\Mcl$.
It is taken to be the image under the so-called {\it c-map} of the moduli space $\KK$
of complexified K\"ahler structure deformations of a CY threefold, i.e. $\Mcl$ is a bundle over $\KK$
constructed in a canonical way \cite{Cecotti:1989qn,Ferrara:1989ik}. The metric on this space is known explicitly
and it has a large isometry group which, in particular, includes the S-duality group $SL(2,\IR)$.
Thus, $\Mcl$ plays the role of a kind of ``symmetric departure point".
However, the deformations break in general all isometries of $\Mcl$ and the only requirement we impose is that they
leave intact the discrete subgroup $SL(2,\IZ)$.

Of course, it is very difficult to encode the QK property and therefore to classify all possible deformations
directly in terms of the metric. Instead, a much more efficient way to do this is to work with the associated twistor spaces
\cite{MR664330}. The twistor space $\cZ$ of a QK manifold $\cM$ is a $\CP$-bundle which carries a canonical complex contact structure
represented by a holomorphic one-form $\cX$. It is then uniquely defined by a set of contact transformations between local trivializations
of this one-form, i.e. local Darboux coordinates for $\cX$.
These transformations are generated by ``holomorphic Hamiltonians" which we call transition functions.
They represent the most compact way to encode the geometry of a QK manifold and its possible deformations.
Our first result in this paper is a transformation property of these functions under the modular group which
ensures that it acts isometrically both on $\cZ$ and $\cM$.

A necessary step to evaluate the metric on $\cM$ starting from transition functions on its twistor space
is to parametrize twistor lines, i.e. to find Darboux coordinates on $\cZ$ as functions
of coordinates on the base $\cM$ and the fiber $\CP$. Generically, this is not possible explicitly,
but it is nevertheless possible to get integral equations determining them.
One of the main results obtained in this paper is a set of such equations written
in terms of coordinates on $\cM$ which transform under $SL(2,\IZ)$
in a simple, actually classical, way. As a result, these equations are manifestly consistent with the modular properties
of the Darboux coordinates and provide the most explicit description
which can be achieved for generic deformations preserving the isometric action of $SL(2,\IZ)$.

In physics terms our construction can be interpreted as follows.
The initial QK manifold $\Mcl$ corresponds to the classical HM moduli space.
Its deformations correspond to the inclusion of quantum corrections which, in particular, contain
the instanton effects due to D-branes and NS5-branes \cite{Becker:1995kb}.
The coordinates on $\cM$ used to parametrize the twistor lines can be viewed as physical fields of type IIB string theory.
As will be clear below, this parametrization appears as a result of a coordinate transformation
which physically yields the non-perturbative mirror map, i.e. a relation between the physical fields
of type IIA and type IIB formulations including all instanton corrections.

In fact, this paper is an extension of the previous work \cite{Alexandrov:2012bu} where the same problem was addressed
under the additional assumption that the allowed deformations preserve two continuous isometries.
In string theory this corresponds to the inclusion of quantum corrections due to D(-1), D1 and D3-instantons, but ignoring
fivebrane instanton effects. This is a consistent approximation since the former and the latter do not mix under S-duality
transformations. However, it is really the latter sector that is the most interesting: whereas D-instanton corrections
to the HM moduli space have been incorporated to all orders \cite{Alexandrov:2008gh,Alexandrov:2009zh},
NS5-brane instantons have been found only in the linear approximation \cite{Alexandrov:2010ca} and represent essentially
the last missing piece in the non-perturbative description of CY compactifications of type II string theory.
Our results provide a general framework to describe these instanton effects. In particular, the transition functions generating them,
which should be an appropriate generalization of the ones found in \cite{Alexandrov:2010ca}, must satisfy the modular
constraint \eqref{master}.

The organization of the paper is as follows.
In section \ref{secQK} we briefly review the twistorial description of QK spaces.
In section \ref{sec-Cmap} we describe the c-map space $\Mcl$, its twistor construction, and the action of the modular group.
Section \ref{sec-Deform} is the core of the paper where we derive our main results.
Section \ref{sec-concl} presents our conclusions.
In appendices one can find details on modular properties of various quantities
and proofs of some modular transformations.

\section{Twistor description of QK spaces}
\label{secQK}

Let us recall that a quaternion-K\"ahler manifold $\cM$ is a $4n$-dimensional Riemannian manifold
whose holonomy group is contained in $USp(n)\times SU(2)$.
It has a triplet of almost complex structures $\vec{J}$ satisfying the quaternionic algebra.
The $J_i$'s are not integrable unless the scalar curvature of $\cM$ vanishes, in which case $\cM$ is
hyperk\"ahler. Nevertheless, it is possible to encode the geometry
of $\cM$ complex analytically, by passing to its twistor space $\cZ$,
the total space of a canonical $\CP$-bundle over $\cM$ where the fiber corresponds to the sphere of almost complex structures.
In contrast to $\cM$, the twistor space has an {\it integrable} complex structure and, moreover, it carries
a canonical {\it complex contact structure}, given by the kernel of the $\cO(2)$-twisted, (1,0)-form
\be
Dt=\text{d}t + p_+ -\I p_3 t +p_- t^2,
\label{Dt}
\ee
where $t$ is a stereographic coordinate on $\CP$ and $(p_{\pm}=-\tfrac12(p_1\mp\I p_2), p_3)$ denotes
the $SU(2)$-part of the Levi-Civita connection on $\cM$.

A more convenient way to encode the complex contact structure is through a holomorphic one-form $\cX$.
More precisely, locally on an open patch $\cU_i\subset \cZ$ there always exists a function $\Phi^{[i]}$
such that the product
\be
\cX^{[i]}=-4\I\, e^{\Phi^{[i]}} Dt/t
\label{cont-oneform}
\ee
is holomorphic, i.e. $\bar\pa$-closed on $\cZ$.
The function $\Phi^{[i]}=\Phi^{[i]}(x^\mu,t)$, known as the ``contact potential",
is holomorphic along the $\CP$ fibers and will play an important role in the following.
In particular, it determines the K\"ahler potential on $\cZ$
\be
\label{Knuflat}
K_{\cZ}\ui{i} = \log\frac{1+\varpi\bar \varpi}{|\varpi|}
+ \Re\Phi\ui{i}(x^\mu,\varpi).
\ee

The advantage of using $\cX$ becomes manifest when one introduces holomorphic Darboux coordinates.
Namely, it is always possible to choose complex coordinates
$(\xi^{\Lambda}_{[i]}, \tilde\xi_\Lambda^{[i]}, \alpha^{[i]})$ in $\cU_i$ such that
the contact one-form \eqref{cont-oneform} takes the canonical form \cite{Neitzke:2007ke,Alexandrov:2008nk}
\be
\cX^{[i]}= \text{d}\alpha^{[i]}+ \xi_{[i]}^\Lambda \text{d}\txi_{\Lambda}^{[i]} .
\label{contact1form}
\ee
Then the contact structure is completely determined by the transformations relating the Darboux coordinate systems
on the overlaps of two patches $\cU_i\cap\cU_j$.
These ``contactomorphisms" must preserve the contact one-form up to a non-vanishing holomorphic factor
\be
\label{glue2}
\CX\ui{i} =  \hat f_{ij}^{2} \, \CX\ui{j}.
\ee
Solving this condition, one finds that such contactomorphisms are all generated by
holomorphic transition functions $\Hij{ij}\in H^1(\cZ, \cO(2))$.
Taking them, as in the case of the usual canonical transformations, to depend on ``initial coordinates" $\xii{i}^\Lambda$
and ``final momenta" $(\txii{j}_\Lambda,\ai{j})$, one obtains the following gluing conditions \cite{Alexandrov:2008nk}
\be
\begin{split}
\xii{j}^\Lambda = &\,  \xii{i}^\Lambda -\p_{\txii{j}_\Lambda }\hHij{ij}
+\xii{j}^\Lambda \, \p_{\ai{j} }\hHij{ij} ,
\\
\txii{j}_\Lambda =&\,  \txii{i}_\Lambda
 + \p_{\xii{i}^\Lambda } \hHij{ij} ,
\\
\ai{j} = &\, \ai{i}
 + \hHij{ij}- \xii{i}^\Lambda \p_{\xii{i}^\Lambda}\hHij{ij} ,
\end{split}
\label{QKgluing}
\ee
which result in the following expression for the coefficients $\hat f_{ij}^2$ appearing in \eqref{glue2}
\be
\hat f_{ij}^2=1-\p_{\ai{j} }\hHij{ij}.
\label{trans_f}
\ee

The gluing conditions \eqref{QKgluing}, supplemented by appropriate regularity conditions,
can be used to find the Darboux coordinates as functions of
coordinates $x^\mu$ on $\cM$ and the coordinate $t$ on $\CP$, which is a necessary step
to compute the metric. In particular, we demand that $\xi^\Lambda$ has simple poles at $t=0$ and $t=\infty$,
whereas all other coordinates are regular.\footnote{In fact, one can also allow for logarithmic singularities
at $t=0$ and $t=\infty$ in $\txi_\Lambda$ and $\alpha$ \cite{Alexandrov:2008nk}. They give rise to the so called
anomalous dimensions, $c_\Lambda$ and $c_\alpha$, the numerical coefficients which supplement the twistor data,
given by the covering and transition functions, defining $\cM$ uniquely. The anomalous dimension $c_\alpha$ plays an important physical
role since it encodes the one-loop $g_s$ correction to the HM metric \cite{Robles-Llana:2006ez,Alexandrov:2007ec,Alexandrov:2008gh}.
But in the type IIB formulation, for which the formalism developed here is supposed to be applied, it is possible to avoid
non-vanishing $c_\alpha$ \cite{Alexandrov:2009qq}.}
Such conditions are particularly convenient for description of the c-map spaces and their deformations
which we are interested in here \cite{Alexandrov:2008nk,Alexandrov:2008gh}.
Using them, the gluing conditions \eqref{QKgluing} can be rewritten as the following integral equations
\bea
\xii{i}^\Lambda(\varpi,x^\mu)& =& A^\Lambda +
\varpi^{-1} Y^\Lambda - \varpi \bY^\Lambda
+\hf\sum_j \oint_{C_j}\frac{\de\varpi'}{2\pi\I \varpi'}\,
\frac{t'+t}{t'-t} \(\p_{\txii{0}_\Lambda }\hHij{j0}
-\xii{0}^\Lambda \, \p_{\ai{0} }\hHij{j0}\) ,
\nonumber \\
\txi_\Lambda^{[i]}(\varpi,x^\mu)& = & B_\Lambda -
\hf \sum_j \oint_{C_j} \frac{\de \varpi'}{2 \pi \I \varpi'} \,
\frac{t'+t}{t'-t}\, \p_{\xii{j}^\Lambda } \hHij{j0} ,
\label{txiqline}
\\
\ai{i}(\varpi,x^\mu)& = & B_\alpha -
\hf \sum_j \oint_{C_j} \frac{\de \varpi'}{2 \pi \I \varpi'} \,
\frac{t'+t}{t'-t}\( \hHij{j0}- \xii{j}^\Lambda \p_{\xii{j}^\Lambda}\hHij{j0}\),
\nonumber
\eea
where $t\in \hU_i$ (the hat denotes the projection to $\CP$), $C_j$ is a contour surrounding $\hU_j$
in the counterclockwise direction, and $\scriptstyle{[0]}$
refers to any patch since the sum over $j$ is independent on its choice.
The complex variables $Y^\Lambda$ and real $ A^\Lambda, B_\Lambda, B_\alpha$ are free parameters playing the role
of coordinates $x^\mu$ on $\cM$. They contain one parameter more than the dimension of $\cM$
because the overall phase rotation of $Y^\Lambda$ can be absorbed by a redefinition
of the $\CP$ coordinate $\varpi$.

Furthermore, combining \eqref{cont-oneform}, \eqref{glue2} and \eqref{trans_f}, one finds
the gluing conditions for the contact potential
\be
\Phi^{[i]}-\Phi^{[j]}  =  \log\(1- \p_{\ai{j} }\hHij{ij}\).
\label{gluPhi}
\ee
Requiring that $\Phi^{[i]}$ is regular in $\cU_i$ and using again \eqref{cont-oneform}, it can be expressed
in terms of solutions of \eqref{txiqline} \cite{Alexandrov:2009zh}
\be
\Phi^{[i]}=\phi+\hf\sum_j \oint_{C_j} \frac{\de \varpi'}{2 \pi \I \varpi'}
\,\frac{t'+t}{t'-t}\, \log\(1- \p_{\ai{0} }\hHij{j0}\),
\label{solcontpot}
\ee
where the first term in \eqref{solcontpot} is a real constant given by
\be
e^\phi=-\frac{\frac{1}{16\pi} \sum_j\oint_{C_j}\frac{\de\varpi}{\varpi}
\(\varpi^{-1} Y^{\Lambda}-\varpi \bY^{\Lambda} \) \p_{\xii{j}^\Lambda } \hHij{j0}}
{\cos\[\frac{1}{4\pi}\sum_j\oint_{C_j}\frac{\de\varpi}{\varpi}\,\log\(1-\p_{\ai{0}}\Hij{j0} \)\]}.
\label{contpotconst}
\ee

It is important to note that all above quantities like the contact one-form, Darboux coordinates, contact potential
are invariant under the combined action of complex conjugation and the antipodal map, $\varsigma[t]=-1/\bar t$, provided
it sends a patch $\cU_i$ to another patch $\cU_{\bi}$, so that the covering is invariant, and the transition function
satisfy
\be
\overline{\varsigma\bigl[\Hij{ij}\bigr]}=\Hij{\bi\bj}.
\ee
These reality conditions are necessary to ensure the reality of the metric which can be obtained once the Darboux coordinates
and the contact potential are explicitly found. The precise procedure is described in \cite{Alexandrov:2008nk,Alexandrov:2008gh}.

Finally, a crucial property of the twistor approach is that any quaternionic isometry of the QK base $\cM$
can be lifted to a {\it holomorphic} isometry on its twistor space \cite{MR872143}. This remains true even for discrete isometries
and allows us to use a powerful combination of holomorphicity and modularity.

\section{C-map spaces and S-duality}
\label{sec-Cmap}

\subsection{The metric and twistorial construction}
\label{subsec-twistcmap}

C-map is a procedure suggested in \cite{Cecotti:1989qn,Ferrara:1989ik}
which associates a QK manifold $\Mcm$ to any (projective) special K\"ahler manifold $\SK$.
Since the latter is determined by a holomorphic prepotential $F(X)$, a function homogeneous of degree 2
dependent of homogeneous complex coordinates $X^\Lambda$ ($\Lambda=0,\dots,n$--1
where $n$--1 is the complex dimension of $\SK$),
the metric on $\Mcm$ is also completely determined by $F(X)$.
It describes a certain bundle over $\SK$ and reads as follows
\be
\begin{split}
\de s_{\Mcm}^2=&\de\phi^2
+4 \de s^2_{\SK}
-\frac{1}{2}\,e^{-\phi}\left(\de\tzeta_\Lambda - \bar\cN_{\Lambda\Sigma} \de\zeta^\Sigma\right)
\Im \cN^{\Lambda\Lambda'}\left(\de\tzeta_{\Lambda'} - \cN_{\Lambda'\Sigma'} \de\zeta^{\Sigma'}\right)
\\
&+ \frac{1}{16}\, e^{-2\phi}\left(\de \sigma + \tzeta_\Lambda \de \zeta^\Lambda -\zeta^\Lambda \de \tzeta_\Lambda\right)^2,
\end{split}
\label{hypmettree}
\ee
where real $\phi,\sigma,\zeta^\Lambda,\tzeta_\Lambda$ and complex $z^a=X^a/X^0$ $(a=1,\dots,n$--1)
compose $4n$ coordinates parametrizing $\Mcm$.
In \eqref{hypmettree}, $\de s^2_{\SK}$ is the metric on $\SK$ with \kahler
potential $\cK=-\log[ \I ( \bar X^\Lambda F_\Lambda - X ^\Lambda \bF_\Lambda)]$,
\be
\label{defcN}
\cN_{\Lambda\Lambda'} =\bF_{\Lambda\Lambda'} +
2\I \frac{ \Im F_{\Lambda\Sigma} X^{\Sigma}
\Im F_{\Lambda'\Sigma'} X^{\Sigma'}}
{X^\Xi \Im F_{\Xi\Xi'}X^{\Xi'}}
\ee
is the Weil period matrix, and we used the notation $F_\Lambda\equiv\p_{X^\Lambda}F$, etc.
The quaternionic structure of the the metric \eqref{hypmettree} has been exhibited in \cite{Ferrara:1989ik}.

Although the above description is very explicit, we need a twistorial construction of $\Mcm$.
Such construction has been suggested in \cite{Alexandrov:2008nk} (on the basis of \cite{Rocek:2005ij,Rocek:2006xb}
where the c-map has been formulated using projective superspace
\cite{Karlhede:1984vr,Hitchin:1986ea,Lindstrom:1987ks}).
Let $\cZ$ be covered by two patches $\cU_+$, $\cU_-$,
which project to open disks centered around the north ($\varpi=0$) and south ($\varpi=\infty$) poles on $\CP$,
and a third patch $\cU_0$ which covers the equator.\footnote{In this paper we are interested only in
the {\it local} metric on $\cM$. Therefore, we do not need to distinguish between the patches covering
the twistor space and their projections to $\CP$, and in what follows the hats denoting such a projection will be omitted.}
The transition functions between complex Darboux coordinates on each patch are defined in terms of the holomorphic prepotential
\be
\label{symp-cmap}
\hHij{+0}= F(\xi) ,
\qquad
\hHij{-0}=\bF(\xi) .
\ee
A simple calculation using \eqref{txiqline} then shows that the Darboux coordinates in the patch $\cU_0$
are given by \cite{Neitzke:2007ke,Alexandrov:2008nk}
\be
\label{gentwi}
\begin{split}
\xii{0}^\Lambda =&\, \zeta^\Lambda +  \varpi^{-1} Y^{\Lambda} -\varpi \,\bY^{\Lambda} ,
\\
\txii{0}_\Lambda =&\, \tzeta_\Lambda + \varpi^{-1} F_\Lambda(Y)-\varpi \,\bF_\Lambda(\bY) ,
\\
\ai{0}=&\,-\hf\bigl( \sigma+\zeta^\Lambda\tzeta_\Lambda\bigr) +\hf\(\bY^\Lambda F_\Lambda(Y)+Y^\Lambda\bF_\Lambda(\bY)\)
\\
&\,
-\zeta^\Lambda\( \varpi^{-1} F_\Lambda(Y)-\varpi \,\bF_\Lambda(\bY) \)
-\( \varpi^{-2} F(Y)+\varpi \,\bF(bY) \),
\end{split}
\ee
where we set
\be
\zeta^\Lambda= A^\Lambda,
\qquad
\tzeta_\Lambda=B_\Lambda+A^\Sigma \Re F_{\Lambda\Sigma} ,
\qquad
\sigma = -2 B_\alpha - A^\Lambda B_\Lambda.
\label{relTAf}
\ee
Furthermore, identifying
\be
Y^\Lambda=2\, e^{(\phi+\cK)/2} X^{\Lambda},
\label{relYphi}
\ee
one can show that the metric derived from the twistor lines \eqref{gentwi} precisely matches the one in \eqref{hypmettree},
whereas the contact potential $\Phi$ computed from \eqref{solcontpot}, \eqref{contpotconst}
is independent on $t$ and coincides with the coordinate $\phi$.

\subsection{S-duality group}
\label{subsec-Sdual}

In this paper we are interested in a specific c-map space, which we call $\Mcl$, and which is obtained from the following prepotential
\be
\label{Flv}
F(X)=-\kappa_{abc} \,\frac{X^a X^b X^c}{6 X^0}.
\ee
The corresponding special K\"ahler manifold coincides with (the large volume limit of)
the moduli space of complexified K\"ahler structure deformations
of a CY threefold, $\SK=\KK$, with $\kappa_{abc}$ being the triple intersection product on $H^2(CY,\IZ)$.
Its c-map image $\Mcl$ is recognized as the classical HM moduli space of type IIB string compactified on the Calabi-Yau.

A particular role of $\Mcl$ becomes transparent by considering its isometries.
Whereas the generic c-map metric \eqref{hypmettree} is invariant under continuous
shifts\footnote{The shifts of $\zeta^\Lambda$ and $\tzeta_\Lambda$ must be supplemented by a coordinate dependent shift of $\sigma$
so that these isometries form the Heisenberg group.}
of $\sigma$, $\zeta^\Lambda$ and $\tzeta_\Lambda$, the metric on $\Mcl$ in addition
carries an isometric action of $SL(2,\IR)$. Its discrete subgroup $SL(2,\IZ)$ is known in physics
literature as {\it S-duality group}. To make this action explicit,
one should perform the following change of variables
\be
\label{symptobd}
\begin{split}
z^a& =b^a+\I t^a,
\qquad
e^\phi = \frac{\tau_2^2}{6}\kappa_{abc}t^a t^b t^c,
\\
\zeta^0&=\tau_1\, ,
\qquad
\zeta^a = - (c^a - \tau_1 b^a)\, ,
\\
\tzeta_a &=  \tc_a+ \frac{1}{2}\, \kappa_{abc} \,b^b (c^c - \tau_1 b^c)\, ,
\qquad
\tzeta_0 =\, \tc_0-\frac{1}{6}\, \kappa_{abc} \,b^a b^b (c^c-\tau_1 b^c)\, ,
\\
\sigma &= -2 (\psi+\frac12  \tau_1 c_0) + \tc_a (c^a - \tau_1 b^a)
-\frac{1}{6}\,\kappa_{abc} \, b^a c^b (c^c - \tau_1 b^c)\, .
\end{split}
\ee
In string theory this change of variables is known as {\it classical mirror map}
and establishes a relation between physical fields of type IIA (on the l.h.s.) and type IIB (on the r.h.s.)
formulations compactified on mirror CY threefolds \cite{Bohm:1999uk}.
Then the metric \eqref{hypmettree} with the prepotential \eqref{Flv}
is invariant under the following transformations
\be\label{SL2Z}
\begin{split}
&\quad \tau \mapsto \frac{a \tau +b}{c \tau + d} \, ,
\qquad
t^a \mapsto t^a |c\tau+d| \, ,
\qquad
\tc_a\mapsto \tc_a \, ,
\\
&
\begin{pmatrix} c^a \\ b^a \end{pmatrix} \mapsto
\begin{pmatrix} a & b \\ c & d  \end{pmatrix}
\begin{pmatrix} c^a \\ b^a \end{pmatrix}\, ,
\qquad
\begin{pmatrix} \tc_0 \\ \psi \end{pmatrix} \mapsto
\begin{pmatrix} d & -c \\ -b & a  \end{pmatrix}
\begin{pmatrix} \tc_0 \\ \psi \end{pmatrix}
\end{split}
\ee
with $ad-bc=1$ and $\tau=\tau_1+\I \tau_2$.

As any quaternionic isometry of a QK manifold, the  $SL(2,\IR)$ action \eqref{SL2Z} can be lifted to the twistor space $\cZ$.
To this end, it should be supplemented by an appropriate action on the $\CP$ fiber.
To write it in the most convenient way, which is directly accessible to generalizations,
let us introduce the zeros of $c\xi^0(t)+d$. This quadratic equation has two roots, $\varpi_\pm^{c,d}$. For example,
in the gauge $X^0=1$, one has $Y^0=\tau_2/2$ and the roots are given
by\footnote{In arbitrary gauge they will still be given by \eqref{poles} multiplied by the phase of $X^0$.}
\be
\varpi_\pm^{c,d} = \frac{ c \tau_1 + d \mp | c\tau + d |}{c \tau_2} ,
\qquad
\varpi^{c,d}_+ \varpi^{c,d}_- = -1.
\label{poles}
\ee
Then the action on the twistor fiber reads as
\be
\label{SL2varpi}
\varpi \mapsto  -\varpi_-^{c,d}\,\frac{\varpi-\varpi_+^{c,d}}{\varpi-\varpi_-^{c,d}},
\ee
and it can be easily checked that, under the combined action of \eqref{SL2Z} and \eqref{SL2varpi},
and after using the classical mirror map \eqref{symptobd},
the Darboux coordinates \eqref{gentwi} indeed transform holomorphically \cite{Alexandrov:2008gh}
\be
\label{SL2Zxi}
\begin{split}
\xi^0 &\mapsto \frac{a \xi^0 +b}{c \xi^0 + d} \, ,
\qquad
\xi^a \mapsto \frac{\xi^a}{c\xi^0+d} \, ,
\qquad
\txi_a \mapsto \txi_a +  \frac{c}{2(c \xi^0+d)} \kappa_{abc} \xi^b \xi^c\, ,
\\
\begin{pmatrix} \txi_0 \\ \alpha \end{pmatrix} &\mapsto
\begin{pmatrix} d & -c \\ -b & a  \end{pmatrix}
\begin{pmatrix} \txi_0 \\ \alpha \end{pmatrix}
+ \frac{1}{6}\, \kappa_{abc} \xi^a\xi^b\xi^c
\begin{pmatrix}
c^2/(c \xi^0+d)\\
-[ c^2 (a\xi^0 + b)+2 c] / (c \xi^0+d)^2
\end{pmatrix} .
\end{split}
\ee

It is now trivial to see that $SL(2,\IR)$ acts isometrically on the twistor space.
Indeed, the transformation \eqref{SL2Zxi} changes the contact one-form \eqref{contact1form} only by an overall holomorphic factor
\be
\hCX\mapsto \hCX/(c\xi^0+d),
\label{SXi}
\ee
thus leaving the complex contact structure invariant. Furthermore, the transformations
\be
\label{SL2phi}
e^\phi \mapsto \frac{e^\phi}{|c\tau+d|}\, ,
\qquad
\frac{|\varpi|}{1+|\varpi|^2} \mapsto \frac{|\varpi|}{1+|\varpi|^2} \frac{|c \xi^0+d|}{|c\tau+d|},
\ee
which can be obtained from \eqref{SL2Z} and \eqref{SL2varpi}, respectively,
ensure that the \kahler potential \eqref{Knuflat} varies by a \kahler transformation, consistent
with the rescaling of $\hCX$,
\be
K_\cZ\mapsto  K_\cZ - \log|c\xi^0+d| \, .
\label{transKZ}
\ee

\section{Modular invariant deformations}
\label{sec-Deform}

\subsection{Twistorial data and modular constraint}
\label{subsec-trfun}

In this section we study deformations of the c-map space $\Mcl$ described above which preserve the discrete
$SL(2,\IZ)$ subgroup of the full isometry group. No other isometries are supposed to survive.
In such situation the twistor framework of section \ref{secQK} appears to be the most efficient tool.
In particular, we can encode all deformations into a refinement of the covering used to define the twistor space of $\Mcl$
and the associated set of transition functions. And the first question we should answer is which conditions
these data should satisfy to preserve the modular invariance?

Roughly speaking, a sufficient set of such conditions requires that the patches refining the covering are mapped into each other
under $SL(2,\IZ)$ and the Darboux coordinates in all these patches transform according to the {\it undeformed} law \eqref{SL2Zxi}.
To make these conditions more explicit, we suggest here a simple twistorial construction,
which realizes the main ideas and allows to avoid cumbersome details.
Later in section \ref{subsec-gen} it will be generalized to more complicated situations.
Although we cannot claim that these generalizations exhaust all possible solutions, they seem to encompass
all physically interesting cases.

Let the twistor space $\cZ$ be defined by the covering
\be
\cZ=\cU_+\cup\cU_-\cup\(\cup_{m,n} \cU_{m,n}\)
\label{coverZ}
\ee
where $\cU_\pm$ cover the north and south poles of $\CP$ as above and $\cU_{m,n}$ are mapped to each other
under the antipodal map and $SL(2,\IZ)$ transformations
\be
\varsigma\[\cU_{m,n}\]= \cU_{-m,-n},
\label{tauU}
\ee
\be
\cU_{m,n}\mapsto \cU_{\hat m,\hat n},
\qquad
\( \hat m\atop \hat n\) =
\(
\begin{array}{cc}
d & -c
\\
-b & a
\end{array}
\)
\( m \atop n \) ,
\label{mappatches}
\ee
including an invariant patch  $\cU_0\equiv \cU_{0,0}$.
Furthermore, we assume in addition that $c\xii{m,n}^0+d$ is non-vanishing in $\cU_{m,n}$ for all $(c,d)\ne (\pm m,\pm n)$,
including $(m,n)=(0,0)$, and has a simple zero for $(c,d)=(m,n)$.\footnote{Note that this assumption
implies that $\cU_{km,kn}$ with $k\ge 1$ are all identical.
In particular, $\cU_{0,\pm k}=\cU_\pm$. However, we will ignore this subtlety and consider all these patches as different.
Otherwise, we would have to pay special attention to coinciding patches and the presentation would become very heavy.
We prefer to sacrifice the rigour in favour of clarity.
A rigorous analysis is also possible and does not change any results.}
Denoting this zero by $\htp^{c,d}$, the reality condition for $\xi^0$ and \eqref{tauU} imply that
\be
\varsigma\bigl[\htp^{c,d}\bigr]=\htp^{-c,-d}\equiv\htm^{c,d}.
\label{realtpm}
\ee

With the covering \eqref{coverZ} we associate the following set of transition functions
\be
\label{transfun}
\Hij{+0}=F(\xii{+}) ,
\qquad
\Hij{-0}=\bF(\xii{-}) ,
\qquad
\Hij{(m,n)0}= G_{m,n}(\xii{m,n},\txii{0},\ai{0}),
\ee
where $F(X)$ is given in \eqref{Flv}. The functions $G_{m,n}$ are not arbitrary. Besides the reality conditions
\be
\overline{\varsigma\bigl[G_{m,n}\bigr]}=G_{-m,-n},
\label{realG}
\ee
they must transform in such a way
so that to ensure the modular invariance. In particular, this implies that the modular transformations should not affect
the contact structure. A simple way to satisfy this condition is to require that the Darboux coordinates transform as
in \eqref{SL2Zxi},
since this leads to the transformation of the contact one-form by a holomorphic factor \eqref{SXi}.
The only new feature here is that the Darboux coordinates in \eqref{SL2Zxi} should be supplemented by patch indices:
due to \eqref{mappatches}, the coordinates in $\cU_{m,n}$ are mapped to those in $\cU_{m',n'}$ with
\be
\( m'\atop n'\) =
\(
\begin{array}{cc}
a & c
\\
b & d
\end{array}
\)
\( m \atop n \).
\label{mpnp}
\ee
Applying now the transformation rule \eqref{SL2Zxi} to the gluing conditions \eqref{QKgluing} between $\cU_0$ and $\cU_{m,n}$,
one easily derives constraints on the transformations of $G_{m,n}$ and its derivatives.
The former reads as\footnote{All variables $\xii{0}^\Lambda$ appearing on the r.h.s. can be expressed through
the variables in the patch $\cU_{m',n'}$ using the gluing conditions. We prefer to present the constraint in the form
\eqref{master} since it is simpler and more convenient for applications. In fact, the proper meaning of this constraint is that
it provides a {\it functional} equation on the transition functions which makes them consistent with the modular action.}
\bea
\begin{split}
G_{m,n}\mapsto &\, \frac{G_{m',n'}}{c\xi^0_{[m',n']}+d}
- \frac{c}{6}\,\kappa_{abc}\, \frac{3 \xii{m',n'}^a - 2 T^a_{m',n'}}{(c\xii{m',n'}^0 +d)(c\xii{0}^0 +d)}\,T^b_{m',n'}T^c_{m',n'}
\\
+ &\,
\frac{c^2}{6}\,\kappa_{abc}\, \frac { 3 \xii{0}^a \xii{m',n'}^b +T^a_{m',n'}T^b_{m',n'}}
{(c\xii{m',n'}^0 +d)(c\xii{0}^0 +d)^2}\, T^c_{m',n'}T^0_{m',n'}
- \frac{c^3}{6}\, \frac{\kappa_{abc}\,\xii{m',n'}^a \xii{m',n'}^b\xii{m',n'}^c  }
{(c\xii{m',n'}^0 +d)^2 (c\xii{0}^0 +d)^2} \,\(T^0_{m',n'}\)^2,
\end{split}
\label{master}
\eea
where we introduced the convenient notation
\be
T_{m,n}^\Lambda=\p_{\txii{0}_\Lambda}G_{m,n}-\xii{0}^\Lambda \p_{\ai{0}} G_{m,n},
\label{defT}
\ee
and the transformation of the derivatives are given in appendix \ref{ap_trder}.
Of course, one can verify the mutual consistency of these transformations.

The constraint \eqref{master} is the main condition ensuring the $SL(2,\IZ)$ invariance of the construction.
Below we prove this by confirming the modular properties of Darboux coordinates and
the simple K\"ahler transformation \eqref{transKZ} of the K\"ahler potential $K_\cZ$.
On the other hand, we do not try to classify solutions to the constraint \eqref{master} and leave this issue for future work
(see, however, some comments in section \ref{sec-concl}).

\subsection{$SL(2,\IZ)$ action on the fiber, critical points and invariant kernel}
\label{subsec-transt}

To verify the holomorphic action of the modular group on the twistor space $\cZ$, we should first of all find
how it is realized on the $\CP$ fiber.
In this respect, we are in a much more complicated situation than in the case with two continuous isometries
analyzed in \cite{Alexandrov:2012bu}. Indeed, in that case it is possible to choose all transition functions
to be independent of two Darboux coordinates, $\alpha$ and $\txi_0$. Then it follows from \eqref{txiqline} that
$\xi^0$ does not get any corrections due to $G_{m,n}$ and is still given by \eqref{gentwi}.
In turn this implies the absence of any modifications in the transformation of the $\CP$ coordinate $t$ \eqref{SL2varpi}.
On the other hand, once all continuous isometries are broken, both $\xi^0$ and the transformation of $t$
are expected to be modified.

Remarkably, it is possible to fix the action of $SL(2,\IZ)$ on the twistor fiber by considering
the transformation of the K\"ahler potential $K_{\cZ}$ and requiring that it coincides with the K\"ahler transformation \eqref{transKZ}.
As shown in appendix \ref{ap-trK}, this condition can be converted to a differential equation which is easily solved.
As a result, the constraint on the transformation of $K_{\cZ}$ boils down to two conditions:
one is a constraint on the transformation of the (shifted) zero mode of the contact potential given in \eqref{transf-hatphi},
whereas the other fixes the transformation of the $\CP$ coordinate to be rational and given by
\be
\glt{c,d}=\Cfm\,\frac{t-\htp^{c,d}}{t-\htm^{c,d}},
\label{transftt}
\ee
where the constant prefactor should satisfy $|\Cfm|=|\htm^{c,d}|$ and we used the notation $\gl{c,d}$
for the $SL(2,\IZ)$ action.
Furthermore, the transformation \eqref{transftt} should satisfy the group law which, in particular, requires that
$\gl{c,d}\[\glt{-c,a}\]=\varpi$. An easy calculation immediately gives that
\be
\Cfpm=\gl{c,d}\[ \htpm^{-c,a}\],
\qquad
\htp^{c,d}\Cfm=\htm^{c,d}\Cfp.
\label{pref-trt}
\ee
In fact, taking into account \eqref{realtpm}, the last condition is equivalent to the above condition on the modulus of $\Cfm$.
Thus, the $SL(2,\IZ)$ action on the twistor fiber is completely expressed in terms of the zeros of $c\xi^0+d$
and is a natural generalization of the classical transformation \eqref{SL2varpi}.\footnote{The prefactor in
\eqref{SL2varpi} agrees with \eqref{pref-trt} since it can be shown that the classical zeros \eqref{poles} satisfy
$\gl{c,d}\[t_\pm^{-c,a}\]=-t_\pm^{c,d}$.}

As will be clear below, not only the zeros of $c\xi^0+d$ are important for our construction, but a crucial role is played also by
the zeros of the derivative $\p_t\xi^0$ or {\it critical points}. On the basis of the undeformed expression for $\xi^0$ \eqref{gentwi},
we expect that there are two such points in the patch $\cU_0$
\be
\p_t\xii{0}^0(\tinvpm)=0,
\label{eqinvpoints}
\ee
which are related by the antipodal map $\varsigma\[\tinvp\]=\tinvm$ (cf. \eqref{realtpm})
and reduce to $\tinvpm=\mp\I$ in the absence of deformations (or for the deformations preserving two continuous isometries
and not modifying $\xi^0$).
A geometric meaning of $\tinvpm$ can be understood by analyzing the modular properties of $\p_t\xi^0$.
One easily finds that $\gl{c,d}[\p_t\xi^0]\sim\p_t\xi^0$, which implies that $\tinvpm$ are {\it invariant} under modular transformations.
It is important to note that $\tinvpm$ should be viewed not just as functions on $\cM$,
but as sections of the bundle $\cZ\to\cM$, and therefore their invariance
is also to be understood as the invariance of sections. Thus, they satisfy the following property
\be
\gl{c,d}[\tinvpm]=\glt{c,d}_{t=\tinvpm}=\Cfm\, \frac{\tinvpm-\htp^{c,d}}{\tinvpm-\htm^{c,d}}.
\label{invpoints}
\ee

Having introduced the critical points, we are now in position to define the so called {\it invariant kernel},
which was the key element for making the modular symmetry explicit
already in the case of two isometries \cite{Alexandrov:2012bu}.
We define it as a $t$-independent shift of the kernel appearing in \eqref{txiqline}, invariant under $SL(2,\IZ)$
after multiplication by the measure:
\be
K(\varpi,\varpi')=\frac{1}{2}\(\frac{\varpi'+\varpi}{\varpi'-\varpi}+\kk(\varpi')\),
\qquad
\frac{\de \varpi'}{\varpi'}\,K(\varpi,\varpi')\mapsto \frac{\de \varpi'}{\varpi'}\,K(\varpi,\varpi').
\label{invkernel2}
\ee
Given the deformed transformation of $t$ \eqref{transftt} and the modular property of the critical points \eqref{invpoints},
it is easy to check that the function
\be
\kk(t)=\frac{\tinvp\tinvm -t^2}{(t-\tinvp)(t-\tinvm)}
\label{reskk}
\ee
does the job and reduces to the known classical result $\frac{1/t-t}{1/t+t}$ in the undeformed case.
As a result, the invariant kernel is found to be\footnote{In fact, the kernel is not the only one possessing
the invariance property. Choosing $\kk^\pm=\frac{\tinvpm+t}{\tinvpm-t}$, one obtains two kernels
$$
\frac{\de t'}{t'}\,K^+(t,t')=\frac{(t-\tinvp)\,\de t'}{(t'-t)(t'-\tinvp)},
\qquad
\frac{\de t'}{t'}\,K^-(t,t')=\frac{(t-\tinvm)\,\de t'}{(t'-t)(t'-\tinvm)}
$$
which are also invariant, so is any their linear combination. In particular, $K=\hf(K^++K^-)$. What distinguishes
this particular combination, or equivalently $\kk(t)$ from \eqref{reskk},
is that it is invariant under the combined action of the antipodal map and complex conjugation
ensuring the reality conditions to be satisfied by various quantities.
In \cite{Alexandrov:2012au} it was argued that the kernels $K^\pm$ can nevertheless be used, at least
in the linear approximation, due to the property
$
\overline{\varsigma\[K^+(t,t')\]}=K^{-}(t,t'),
$
and appear to be useful in the context of the so called large volume limit, where $t^a\to\infty$ keeping other coordinates fixed.
In this limit all integrals can be evaluated by the saddle point approximation with the saddle points
approaching the critical points $\tinvpm$.
The problem is that the formulation based on the single kernel $K$ leads to appearance
of terms diverging in the large volume limit which however can be removed by a coordinate redefinition.
The passage to the kernels $K^\pm$ takes care about this redefinition and directly provides final results.
Unfortunately, we do not know how to extend these results to the non-perturbative treatment of generic deformations
which we are doing here.}
\be
\begin{split}
\frac{\de t'}{t'}\,K(t,t')=&\,
\frac{\de t'}{2}\, \frac{(t-\tinvp)(t'-\tinvm)+(t-\tinvm)(t'-\tinvp)}{(t'-t)(t'-\tinvp)(t'-\tinvm)} .
\end{split}
\label{fullinvkernel}
\ee
Together with the critical points, it represents the main building block of the modular invariant construction.

\subsection{Twistor lines and ``mirror map"}
\label{subsec-mirmap}

Let us now consider the Darboux coordinates corresponding to the twistor data of section \ref{subsec-trfun} and prove that they transform
according to the classical law \eqref{SL2Zxi} provided the transition functions satisfy the modular constraint \eqref{master}.
To this end, one has to express the coordinates $Y^\Lambda, A^\Lambda,B_\Lambda,B_\alpha$, parametrizing twistor lines in \eqref{txiqline},
in terms of the coordinates $\tau,b^a,t^a,c^a,\tc_a,\tc_0,\psi$, having simple modular transformation properties \eqref{SL2Z},
in such a way that the integral equations determining twistor lines stay invariant under $SL(2,\IZ)$ transformations.
Essentially, such change of coordinates would provide a generalization of the classical mirror map \eqref{symptobd}
since the type IIA physical fields typically have very simple relations to the former set of coordinates fixed by
the symplectic invariance of the type IIA construction.
For example, for the deformations corresponding to D-instanton corrections they have been found in \cite[Eq. (3.10)]{Alexandrov:2009zh}.
The situation can be illustrated by the following diagram

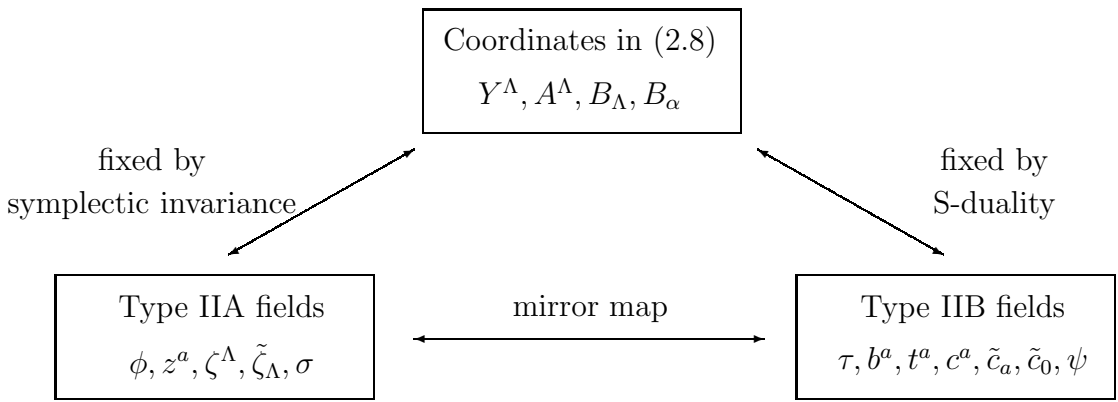
\unitlength 0.7mm 
\linethickness{0.4pt}
\ifx\plotpoint\undefined\newsavebox{\plotpoint}\fi 
\begin{picture}(261,85)(50,58)
\put(159.5,130.75){\makebox(0,0)[cc]{Coordinates in \eqref{txiqline}}}
\put(159.5,120.75){\makebox(0,0)[cc]{$Y^\Lambda, A^\Lambda,B_\Lambda,B_\alpha$}}
\put(129.75,113.5){\framebox(60.25,23.25)[ct]{}}
\put(60,63){\framebox(60.25,23.25)[ct]{}}
\put(200.75,63){\framebox(60.25,23.25)[ct]{}}
\put(91.5,80){\makebox(0,0)[cc]{Type IIA fields}}
\put(91.5,70){\makebox(0,0)[cc]{$\phi,z^a,\zeta^\Lambda,\tzeta_\Lambda,\sigma$}}
\put(232.25,80){\makebox(0,0)[cc]{Type IIB fields}}
\put(232.25,70){\makebox(0,0)[cc]{$\tau,b^a,t^a,c^a,\tc_a,\tc_0,\psi$}}
\put(194.5,74.25){\vector(1,0){.07}}\put(128,74.25){\vector(-1,0){.07}}\put(128,74.25){\line(1,0){66.5}}
\put(92.75,90.25){\vector(-2,-1){.07}}\put(128,110.25){\vector(2,1){.07}}\multiput(128,110.25)(-.0594435076,-.0337268128){593}{\line(-1,0){.0594435076}}
\put(228.5,90.25){\vector(2,-1){.07}}\put(193.25,110.25){\vector(-2,1){.07}}\multiput(193.25,110.25)(.0594435076,-.0337268128){593}{\line(1,0){.0594435076}}
\put(161.5,80){\makebox(0,0)[cc]{mirror map}}
\put(238.25,103.75){\makebox(0,0)[cc]{$\displaystyle{ \text{fixed by} \atop \text{S-duality}}$}}
\put(78.25,103.75){\makebox(0,0)[cc]{$\displaystyle{ \text{fixed by} \atop \text{symplectic invariance}}$}}
\end{picture}

For simplicity, in the following we will call the change of coordinates given by the right arrow also by ``mirror map".
As we will see, for the zero modes $A^\Lambda,B_\Lambda,B_\alpha$ it is given essentially
by the shift converting the integration kernel into the invariant one \eqref{invkernel2}.
The main non-trivial problem is to fix the mirror map for the variables $Y^\Lambda$.
It is provided by the following Lemma whose proof can be found in appendix \ref{ap_Lemma}.

\smallskip

{\bf Lemma:} {\it Let us define\footnote{The prime on the sum indicates that it runs over $(m,n)\in \IZ^2/{(0,0)}$.}
\be
Y^\Lambda=
 \I\tau_2\,\frac{\tinvp\tinvm}{\tinvm-\tinvp}\,\cY^\Lambda+
\frac{\tinvp\tinvm}{2}{\sum_{m,n}}' \oint_{C_{m,n}}\frac{\de \varpi}{2 \pi \I}\,
\frac{ \tinvp(\tinvm-t)+\tinvm(\tinvp-t)}{(t-\tinvp)^2(t-\tinvm)^2}\,T^\Lambda_{m,n},
\label{defYL}
\ee
where
\be
\cY^0=1,
\qquad
\cY^a =
\frac{b^a\(\tinvm e^{\I U}-\tinvp e^{-\I U}\)-2\sqrt{\tinvp \tinvm}\,t^a}{(\tinvm-\tinvp)\cos U}
+ \I\tau_2^{-1}\tan U \,\cI^a
\ee
and
\bea
\cI^a &=& \frac{(\tinvm-\tinvp)^2}{4\pi}{\sum_{m,n}}' \oint_{C_{m,n}}\de \varpi\,
\frac{\varpi \,T^a_{m,n}}{(\varpi-\tinvp)^2 (\varpi-\tinvm)^2},
\label{defI}
\\
U &=&\frac{\tinvm-\tinvp}{4\pi}{\sum_{m,n}}' \oint_{C_{m,n}}\de \varpi\,
\frac{ \log\(1- \p_{\ai{0} }G_{m,n}\)}{(\varpi -\tinvp) (\varpi- \tinvm)}.
\label{defU}
\eea
Then, provided at the critical points one has
\be
\begin{array}{rclcrcl}
\xii{0}^0(\tinvp)&=&\tau,
& \qquad &
\xii{0}^0(\tinvm)&=&\bar\tau,
\\
\xii{0}^a (\tinvp) &=& -c^a + \tau b^a,
& \qquad &
\xii{0}^a (\tinvm) &=& -c^a + \bar \tau b^a,
\end{array}
\label{cond-spm}
\ee
\be
\tinvpm\p_t\xii{0}^a (\tinvpm) = \pm\frac{2\I\tau_2 t^a}{\cos U}\, \frac{ \sqrt{\tinvp\tinvm}}{\tinvm-\tinvp}-\I e^{\pm \I U}L^a,
\label{cond-spmder}
\ee
where $L^a$ is an arbitrary real vector,
and using the notation $\rhoi{c,d}$ from \eqref{defrho},
the quantities \eqref{defYL} transform as
\be
Y^0 \ \mapsto\
\frac{\Cfm\htp^{c,d}}{c(\htp^{c,d} - \htm^{c,d})^2 \rhoi{c,d}(\htp^{c,d})},
\qquad
Y^a \ \mapsto\
- \frac{\Cfm\htp^{c,d}\xii{c,d}^a(\htp^{c,d})}{(\htp^{c,d} - \htm^{c,d})^2 \rhoi{c,d}(\htp^{c,d})}.
\label{transfYL}
\ee
}

\smallskip

This Lemma makes it easy to prove the Theorem which represents one of the main results of this work.

\smallskip

{\bf Theorem:} {\it Provided $Y^\Lambda$ is given by \eqref{defYL} and the zero modes are given by
the following expressions
\bea
A^\Lambda &=& \zeta^\Lambda_{\rm cl}
-\hf\, (\tinvp+\tinvm)\(\frac{Y^\Lambda}{\tinvp\tinvm}-\bY^\Lambda\)
+\hf\sum_j \oint_{C_j}\frac{\de\varpi}{2\pi\I \varpi}\,
\kk(\varpi)\(\p_{\txii{0}_\Lambda }\hHij{j0}-\xii{0}^\Lambda \, \p_{\ai{0} }\hHij{j0}\) ,
\nn
\\
B_\Lambda &=& \tzeta_\Lambda^{\rm cl}
-\hf\sum_j \oint_{C_j}\frac{\de\varpi}{2\pi\I \varpi}\,
\kk(\varpi)\p_{\xii{j}^\Lambda}\Hij{j0},
\label{qmirror}
\\
B_\alpha &=& (\sigma+\zeta^\Lambda \tzeta_\Lambda)^{\rm cl}
-\hf\sum_j \oint_{C_j}\frac{\de\varpi}{2\pi\I \varpi}\,
\kk(\varpi)\( \hHij{j0}- \xii{j}^\Lambda \p_{\xii{j}^\Lambda}\hHij{j0}\),
\nn
\eea
where the index {\rm\small cl} refers to the quantities given by the classical mirror map $\eqref{symptobd}$,
the integral equations \eqref{txiqline} are invariant under modular transformations and the Darboux coordinates transform as
in \eqref{SL2Zxi}.
}

The proof of this Theorem is given in appendix \ref{ap_Theorem}. Here we would like just to concentrate
on the Darboux coordinate $\xi^0$ which plays the role of the modular parameter on the twistor space.
Substituting the mirror map for $Y^0$ and $A^0$, it is found to be
\bea
\begin{split}
\xii{0}^0(t)=&\,
\tau_1+\frac{\I\tau_2}{\tinvm-\tinvp}\(\frac{\tinvp\tinvm}{t}-t\)
\\
&\,
+ \frac{(t-\tinvp)(t-\tinvm)}{2t}\, {\sum_{m,n}}' \oint_{C_{m,n}}\frac{\de \varpi'}{2 \pi \I}\,t'\,
\frac{(t-\tinvp)(t'-\tinvm)+(t-\tinvm)(t'-\tinvp)}{(t'-t)(t'-\tinvp)^2(t'-\tinvm)^2}\,T^0_{m,n}.
\end{split}
\label{resxi0}
\eea
As a result, it trivially satisfies the condition \eqref{cond-spm} of the Lemma.
Furthermore, it is now possible to write down explicitly the equation \eqref{eqinvpoints} for the critical points
which takes the following form
\be
\frac{\tinvp+\tinvm}{(\tinvp-\tinvm)^3}
=\frac{1}{4\pi\tau_2}\, {\sum_{m,n}}' \oint_{C_{m,n}}\de \varpi \,
\frac{\varpi\,T^0_{m,n}}{(\varpi -\tinvp)^2 (\varpi- \tinvm)^2} .
\label{eqspm}
\ee
Supplemented by the condition $\varsigma\[\tinvp\]=\tinvm$,
it can be solved by perturbative expansion in powers of (integrals of) the transition functions $T^0_{m,n}$ generating deformations.

Although we have shown the correct transformations of Darboux coordinates under the combined action of \eqref{SL2Z} and \eqref{transftt},
a piece of the proof is still missing. Namely, we saw in section \ref{subsec-transt}
that the transformation of the fiber coordinate belongs to $SL(2,\IZ)$
only if $|\Cfm|=|\htm^{c,d}|$, which is equivalent to the second relation in \eqref{pref-trt}.
Given the explicit expression for $\xi^0$, we can now prove that it indeed holds.
To this end, let us apply an $SL(2,\IZ)$ transformation to the identity $-c\xii{-c,a}^0(\htp^{-c,a})+a=0$.
Using \eqref{resxi0}, one can demonstrate that the transformation of the l.h.s. turns out to be proportional to
\be
\[ c\xii{0,1}^0\( \frac{\cE}{\Cfm-\Cfp}\)+d\]^{-1},
\qquad \mbox{where}\quad \cE=\htp^{c,d}\Cfm-\htm^{c,d}\Cfp.
\ee
Since this expression should vanish, $\frac{\cE}{\Cfm-\Cfp}$ should be a singular point of $\xi^0$ and therefore coincide with
the north pole of $\CP$. Thus, one concludes that $\cE=0$ as required.

Finally, note that the whole construction is invariant under $U(1)$ phase rotations
\be
t\ \mapsto\ e^{\I\varphi} t,
\qquad
\tinvpm\ \mapsto e^{\I\varphi}\tinvpm.
\ee
They induce a similar rotation of the complex variables $Y^\Lambda \mapsto e^{\I\varphi} Y^\Lambda$,
so that this symmetry coincides with the one mentioned below \eqref{txiqline} and responsible for removing
one auxiliary coordinate. In the given case this additional coordinate is provided by the phase
of the critical points, since it is not determined by the defining equation \eqref{eqspm}.
A convenient way to fix it, for example, is to require that the critical points $\tinvpm$ are pure imaginary.

\subsection{Modular invariant potential}
\label{subsec-contpot}

To complete the construction, it remains to prove the last remaining transformation, which is the modular transformation \eqref{transf-hatphi}
of the zero mode of the contact potential. In appendix \ref{ap-trK} it appears as a necessary condition
for the K\"ahler potential $K_\cZ$ to have the right modular properties.
Substituting $\kk(\varpi)$ \eqref{reskk} and evaluating the integral as in \eqref{evalint}, this condition can be written
more explicitly as
\be
\gl{c,d}[\hat\phi]=\hat\phi
+\hf\log \frac{(\tinvp-\htp^{c,d})(\tinvm-\htm^{c,d})(\tinvp-\htm^{c,d})(\tinvm-\htp^{c,d})}{\tinvp\tinvm(\htm^{c,d}-\htp^{c,d})^2|c\tau+d|^2},
\label{tran-hatphi}
\ee
where the shifted zero mode $\hat\phi$ is defined in \eqref{defhatphi}.
Furthermore, from \eqref{invpoints} it is easy to see that the inhomogeneous term on the r.h.s. is equal to
$\log\frac{\tau_2^{-1/2}|\tinvp-\tinvm|}{\gl{c,d}[\tau_2^{-1/2}|\tinvp-\tinvm|]}$.\footnote{Note that the property $\varsigma\[\tinvp\]=\tinvm$
ensures that $|\tinvp-\tinvm|^2=-\frac{(\tinvp-\tinvm)^2}{\tinvp\tinvm}$. }
As a result, the transformation \eqref{tran-hatphi} is equivalent to the requirement that the following function on $\cM$
\bea
\label{constpartcpchk}
\begin{split}
\Fi =&\,\frac{|\tinvp-\tinvm|\, e^{-\frac12{\sum\limits_{m,n}}'\oint_{C_{m,n}}\frac{\de \varpi}{2 \pi \I \varpi}
\,\frac{\tinvp\tinvm -t^2}{(t-\tinvp)(t-\tinvm)} \log\(1- \p_{\ai{0} }G_{m,n}\)   }}
{\tau_2^{1/2}\cos\[ \frac{1}{4\pi} {\sum\limits_{m,n}}'\oint_{C_{m,n}}\frac{\de\varpi}{\varpi}\,\log\(1-\p_{\ai{0}}G_{m,n} \)\]}
\Biggl\{\hf\,\Im \( Y^\Lambda \bF_\Lambda(\bY)\)
\Biggr.
\\
& \left.
-\frac{1}{16\pi}{\sum_{m,n}}' \oint_{C_{m,n}}\frac{\de\varpi}{\varpi}\[
\(\varpi^{-1} Y^{\Lambda}-\varpi \bY^{\Lambda} \)\p_{\xii{m,n}^\Lambda}G_{m,n}
+\(\varpi^{-1} F_{\Lambda}(Y)-\varpi \bF_\Lambda(\bY) \)T_{m,n}^\Lambda\]\right\}
\end{split}
\eea
is modular invariant. Here we evaluated the constant part of the contact potential given in \eqref{contpotconst} for our twistor data
and took into account the additional exponential factor coming from the difference between $\phi$ and $\hat\phi$ \eqref{defhatphi}
and originating from the shift of the kernel done in \eqref{Kahlpot}.
The invariance of the potential \eqref{constpartcpchk} is verified in appendix \ref{ap_zeromode}.

This completes the proof that the twistor space $\cZ$ defined by the transition functions \eqref{transfun}
and, as a consequence, the QK manifold $\cM$ both carry
an isometric action of the modular group $SL(2,\IZ)$.
Besides, we have found a modular invariant parametrization of the twistor lines provided by the coordinate change
\eqref{defYL} and \eqref{qmirror} (see \eqref{qmirrorG} for their explicit form).
And finally we have constructed a modular invariant function \eqref{constpartcpchk}
on the QK base, which can be considered as a non-trivial potential characterizing the deformations.

\subsection{Generalizations}
\label{subsec-gen}

Although the conditions on the twistor data spelled out in section \ref{subsec-trfun} are sufficient to ensure
the isometric action of $SL(2,\IZ)$, they are not necessary conditions. One can easily imagine more complicated coverings
and the associated sets of transition functions for which our construction still goes through.
Here we would like to present several such generalizations which, as we believe, encompass all physically interesting cases.

\subsubsection{Regular terms}

The first possible generalization is to slightly relax the modular constraint \eqref{master} on the transition functions.
Namely, a modular transformation can result in the appearance of terms regular in the target patch $\cU_{m',n'}$
\be
G_{m,n}
\mapsto \frac{G_{m',n'}}{c\xii{m',n'}^0+d}
\ +\mbox{non-linear terms}\
+\ \mbox{regular terms}\, .
\label{StransGreg}
\ee
This is possible because the additional regular terms can be removed by a local contactomorphism performed in the patch $\cU_{m',n'}$
and represent an inherent ambiguity of the twistor framework. In our language their contributions disappear
as a result of contour integration, which is especially easy to see if they
depend only on $\xii{m',n'}^\Lambda$.

Such regular terms do show up in the case of transition functions generating D(-1)-D1-instanton
corrections to the HM moduli space \cite{Alexandrov:2012bu}
so that the generalization \eqref{StransGreg} cannot be ignored.

\subsubsection{Refining the patches}

Another evident generalization is to further refine the patch $\cU_0$ and represent it as a union of sets invariant under
$SL(2,\IZ)$. Thus, the covering of the twistor space in the most general situation will look like
\be
\cZ=\cU_+\cup\cU_-\cup\cU_{0}\cup\Bigl[ \mathop{\cup}\limits_s\Bigl({\mathop{\cup}\limits_{\vec I_s}}\ \cU_{\vec I_s}^s\Bigr)\Bigr],
\qquad
\cU_{\vec I_s}^s\ \mathop{\mapsto}\limits^{\gl{} }\ \cU_{\gl{}^{\rm tr} \vec I_s}^s,
\label{gencover}
\ee
where the index $s$ labels the invariant sets and the patches in each set are enumerated by vectors $\vec I_s$
belonging to some finite dimensional representations of $SL(2,\IZ)$.
The transition functions $\Hij{(s;\vec I_s)0}= G_{\vec I_s}^s$ should satisfy the same modular constraint \eqref{master}
(or its generalization \eqref{StransGreg}) with the indices $(m',n')$ replaced by $\(s,(\gl{}^{\rm tr})^{-1}\vec I_s\)$.
In this picture all invariant sets appear on equal footing except that we require that only one of them,
corresponding to a fundamental representation,
contains zeros of $c\xi^0+d$, whereas in all other patches this combination is non-vanishing.

\subsubsection{Open contours}

Although the picture presented above already provides an exhaustive generalization, in practice
one encounters some twistor descriptions which do not fit this picture.
The point is that transition functions can have branch cuts.
They require special attention, but typically it is still possible to encode the Darboux coordinates
for the contact one-form by means of equations of the type \eqref{txiqline} with contours $C_j$ which now may be more complicated
than just surrounding the patches.
A typical example is a ``figure-eight" contour going around branch points of a logarithmic cut
(see, for instance, \cite[Section 3.4]{Alexandrov:2008ds}).
Nevertheless, since the contours remain closed, our results on modular invariance still hold
because all that we need is to be able to evaluate various contour integrals by residues.

However, it is often possible to trade the description in terms of transition functions with branch cuts
and closed contours for a description based on {\it open contours} joining branch points of the original framework.
The transition functions associated to these open contours are then given by discontinuities of the original functions.
The new description is usually much more economic and elegant and may be indispensable for getting important insights.
For instance, the twistor description of D-instanton corrections to CY compactifications of type II string theory
has been found in terms of open contours joining the north and south poles of $\CP$ (called {\it BPS rays}) \cite{Alexandrov:2008gh},
and it was crucial for understanding the relation of this description to the wall-crossing phenomenon
\cite{ks,Gaiotto:2008cd,Alexandrov:2011ac}.

Taking into account this possibility, the most general twistorial data defining (locally)
a QK manifold are given by a set of contours (closed or open)
on $\CP$ and transition functions generating contact transformations
through each of these contours. A natural question then is whether our results on modularity
can be extended to include the cases where some of the contours are open?

At first sight this seems to be unlikely since one cannot use anymore the evaluation of integrals by residues,
which was extensively done to prove our results. On the other hand, all these results
can be verified perturbatively by expanding in powers of $G_{m,n}$.\footnote{Of course, for generic deformations
breaking all continuous isometries the expansion is infinite. We have performed the explicit perturbative check
of modular invariance only to the second order.
For the deformations preserving two continuous isometries, however, the expansion stops at third order \cite{Alexandrov:2012bu}.}
Such computation does not require explicit evaluation of integrals and does not seem to depend on the nature of contours,
thereby suggesting that all the results continue to hold for the case of open contours as well.

\begin{figure}
\unitlength .7mm 
\linethickness{0.4pt}
\ifx\plotpoint\undefined\newsavebox{\plotpoint}\fi 
\begin{picture}(196.25,135)(0,42)
\put(160,75){\circle*{2.236}}
\put(193.214,75){\line(0,1){1.2845}}
\put(193.189,76.284){\line(0,1){1.2826}}
\multiput(193.114,77.567)(-.04135,.42624){3}{\line(0,1){.42624}}
\multiput(192.99,78.846)(-.04335,.31824){4}{\line(0,1){.31824}}
\multiput(192.817,80.119)(-.044502,.253061){5}{\line(0,1){.253061}}
\multiput(192.594,81.384)(-.045213,.209292){6}{\line(0,1){.209292}}
\multiput(192.323,82.64)(-.045663,.177761){7}{\line(0,1){.177761}}
\multiput(192.003,83.884)(-.04594,.153879){8}{\line(0,1){.153879}}
\multiput(191.636,85.115)(-.046095,.1351){9}{\line(0,1){.1351}}
\multiput(191.221,86.331)(-.046157,.119894){10}{\line(0,1){.119894}}
\multiput(190.759,87.53)(-.050759,.11802){10}{\line(0,1){.11802}}
\multiput(190.252,88.71)(-.050259,.105426){11}{\line(0,1){.105426}}
\multiput(189.699,89.87)(-.049774,.094786){12}{\line(0,1){.094786}}
\multiput(189.102,91.007)(-.0492945,.0856527){13}{\line(0,1){.0856527}}
\multiput(188.461,92.121)(-.0488151,.0777049){14}{\line(0,1){.0777049}}
\multiput(187.777,93.209)(-.0483315,.0707084){15}{\line(0,1){.0707084}}
\multiput(187.052,94.269)(-.0510299,.0687863){15}{\line(0,1){.0687863}}
\multiput(186.287,95.301)(-.0502987,.0625888){16}{\line(0,1){.0625888}}
\multiput(185.482,96.302)(-.0495826,.0570322){17}{\line(0,1){.0570322}}
\multiput(184.639,97.272)(-.0488761,.0520124){18}{\line(0,1){.0520124}}
\multiput(183.759,98.208)(-.0508511,.0500833){18}{\line(-1,0){.0508511}}
\multiput(182.844,99.11)(-.0558528,.0509075){17}{\line(-1,0){.0558528}}
\multiput(181.895,99.975)(-.0577798,.0487094){17}{\line(-1,0){.0577798}}
\multiput(180.912,100.803)(-.0633466,.0493408){16}{\line(-1,0){.0633466}}
\multiput(179.899,101.593)(-.0695546,.0499776){15}{\line(-1,0){.0695546}}
\multiput(178.856,102.342)(-.0765379,.0506253){14}{\line(-1,0){.0765379}}
\multiput(177.784,103.051)(-.0784385,.0476275){14}{\line(-1,0){.0784385}}
\multiput(176.686,103.718)(-.0863926,.0479859){13}{\line(-1,0){.0863926}}
\multiput(175.563,104.342)(-.095532,.048326){12}{\line(-1,0){.095532}}
\multiput(174.416,104.922)(-.106178,.04865){11}{\line(-1,0){.106178}}
\multiput(173.248,105.457)(-.118778,.048958){10}{\line(-1,0){.118778}}
\multiput(172.061,105.946)(-.133981,.049253){9}{\line(-1,0){.133981}}
\multiput(170.855,106.39)(-.152758,.049539){8}{\line(-1,0){.152758}}
\multiput(169.633,106.786)(-.17664,.049822){7}{\line(-1,0){.17664}}
\multiput(168.396,107.135)(-.208174,.050112){6}{\line(-1,0){.208174}}
\multiput(167.147,107.435)(-.251947,.050429){5}{\line(-1,0){.251947}}
\multiput(165.887,107.688)(-.31714,.05081){4}{\line(-1,0){.31714}}
\multiput(164.619,107.891)(-.31886,.03851){4}{\line(-1,0){.31886}}
\put(163.343,108.045){\line(-1,0){1.2805}}
\put(162.063,108.149){\line(-1,0){1.2835}}
\put(160.779,108.204){\line(-1,0){1.2847}}
\put(159.495,108.21){\line(-1,0){1.284}}
\put(158.211,108.165){\line(-1,0){1.2813}}
\multiput(156.93,108.071)(-.42556,-.04783){3}{\line(-1,0){.42556}}
\multiput(155.653,107.928)(-.31755,-.04819){4}{\line(-1,0){.31755}}
\multiput(154.383,107.735)(-.252355,-.048346){5}{\line(-1,0){.252355}}
\multiput(153.121,107.493)(-.20858,-.048391){6}{\line(-1,0){.20858}}
\multiput(151.869,107.203)(-.177045,-.048361){7}{\line(-1,0){.177045}}
\multiput(150.63,106.864)(-.153162,-.048276){8}{\line(-1,0){.153162}}
\multiput(149.405,106.478)(-.134383,-.048145){9}{\line(-1,0){.134383}}
\multiput(148.195,106.045)(-.119178,-.047975){10}{\line(-1,0){.119178}}
\multiput(147.004,105.565)(-.106576,-.047771){11}{\line(-1,0){.106576}}
\multiput(145.831,105.04)(-.095928,-.047536){12}{\line(-1,0){.095928}}
\multiput(144.68,104.469)(-.086786,-.0472708){13}{\line(-1,0){.086786}}
\multiput(143.552,103.855)(-.0848929,-.0505917){13}{\line(-1,0){.0848929}}
\multiput(142.448,103.197)(-.0769534,-.0499915){14}{\line(-1,0){.0769534}}
\multiput(141.371,102.497)(-.069965,-.0494015){15}{\line(-1,0){.069965}}
\multiput(140.321,101.756)(-.063752,-.0488159){16}{\line(-1,0){.063752}}
\multiput(139.301,100.975)(-.0581801,-.0482305){17}{\line(-1,0){.0581801}}
\multiput(138.312,100.155)(-.0562714,-.0504445){17}{\line(-1,0){.0562714}}
\multiput(137.356,99.298)(-.0512629,-.0496617){18}{\line(-1,0){.0512629}}
\multiput(136.433,98.404)(-.049304,-.051607){18}{\line(0,-1){.051607}}
\multiput(135.546,97.475)(-.050052,-.0566208){17}{\line(0,-1){.0566208}}
\multiput(134.695,96.512)(-.0508139,-.0621712){16}{\line(0,-1){.0621712}}
\multiput(133.882,95.517)(-.0483715,-.0640899){16}{\line(0,-1){.0640899}}
\multiput(133.108,94.492)(-.0489138,-.0703068){15}{\line(0,-1){.0703068}}
\multiput(132.374,93.437)(-.0494552,-.0772991){14}{\line(0,-1){.0772991}}
\multiput(131.682,92.355)(-.0500002,-.0852427){13}{\line(0,-1){.0852427}}
\multiput(131.032,91.247)(-.050555,-.094372){12}{\line(0,-1){.094372}}
\multiput(130.425,90.115)(-.046868,-.096257){12}{\line(0,-1){.096257}}
\multiput(129.863,88.96)(-.047029,-.106906){11}{\line(0,-1){.106906}}
\multiput(129.345,87.784)(-.047145,-.119509){10}{\line(0,-1){.119509}}
\multiput(128.874,86.589)(-.047209,-.134714){9}{\line(0,-1){.134714}}
\multiput(128.449,85.376)(-.047209,-.153494){8}{\line(0,-1){.153494}}
\multiput(128.071,84.148)(-.047129,-.177377){7}{\line(0,-1){.177377}}
\multiput(127.741,82.906)(-.04694,-.208912){6}{\line(0,-1){.208912}}
\multiput(127.46,81.653)(-.04659,-.252685){5}{\line(0,-1){.252685}}
\multiput(127.227,80.39)(-.04598,-.31787){4}{\line(0,-1){.31787}}
\multiput(127.043,79.118)(-.04487,-.42588){3}{\line(0,-1){.42588}}
\put(126.908,77.84){\line(0,-1){1.2819}}
\put(126.823,76.559){\line(0,-1){1.2842}}
\put(126.788,75.274){\line(0,-1){1.2846}}
\put(126.802,73.99){\line(0,-1){1.2831}}
\put(126.866,72.707){\line(0,-1){1.2797}}
\multiput(126.979,71.427)(.04072,-.31859){4}{\line(0,-1){.31859}}
\multiput(127.142,70.152)(.04241,-.25342){5}{\line(0,-1){.25342}}
\multiput(127.354,68.885)(.043483,-.209659){6}{\line(0,-1){.209659}}
\multiput(127.615,67.627)(.044193,-.178132){7}{\line(0,-1){.178132}}
\multiput(127.924,66.38)(.051049,-.176289){7}{\line(0,-1){.176289}}
\multiput(128.282,65.146)(.0506,-.15241){8}{\line(0,-1){.15241}}
\multiput(128.687,63.927)(.050183,-.133635){9}{\line(0,-1){.133635}}
\multiput(129.138,62.724)(.049782,-.118435){10}{\line(0,-1){.118435}}
\multiput(129.636,61.54)(.049387,-.105837){11}{\line(0,-1){.105837}}
\multiput(130.179,60.376)(.048989,-.095194){12}{\line(0,-1){.095194}}
\multiput(130.767,59.234)(.0485855,-.0860569){13}{\line(0,-1){.0860569}}
\multiput(131.399,58.115)(.0481717,-.0781054){14}{\line(0,-1){.0781054}}
\multiput(132.073,57.021)(.0477459,-.0711051){15}{\line(0,-1){.0711051}}
\multiput(132.789,55.955)(.05046,-.0692054){15}{\line(0,-1){.0692054}}
\multiput(133.546,54.917)(.04978,-.063002){16}{\line(0,-1){.063002}}
\multiput(134.343,53.909)(.0491099,-.0574397){17}{\line(0,-1){.0574397}}
\multiput(135.178,52.932)(.0484449,-.0524143){18}{\line(0,-1){.0524143}}
\multiput(136.05,51.989)(.0504357,-.0505016){18}{\line(0,-1){.0505016}}
\multiput(136.958,51.08)(.052351,-.0485133){18}{\line(1,0){.052351}}
\multiput(137.9,50.206)(.0573756,-.0491849){17}{\line(1,0){.0573756}}
\multiput(138.875,49.37)(.062937,-.0498623){16}{\line(1,0){.062937}}
\multiput(139.882,48.573)(.0691395,-.0505504){15}{\line(1,0){.0691395}}
\multiput(140.919,47.814)(.0710427,-.0478387){15}{\line(1,0){.0710427}}
\multiput(141.985,47.097)(.0780425,-.0482737){14}{\line(1,0){.0780425}}
\multiput(143.078,46.421)(.0859934,-.0486978){13}{\line(1,0){.0859934}}
\multiput(144.195,45.788)(.09513,-.049114){12}{\line(1,0){.09513}}
\multiput(145.337,45.198)(.105773,-.049525){11}{\line(1,0){.105773}}
\multiput(146.5,44.654)(.11837,-.049937){10}{\line(1,0){.11837}}
\multiput(147.684,44.154)(.133569,-.050358){9}{\line(1,0){.133569}}
\multiput(148.886,43.701)(.152344,-.050799){8}{\line(1,0){.152344}}
\multiput(150.105,43.295)(.154195,-.044869){8}{\line(1,0){.154195}}
\multiput(151.339,42.936)(.178074,-.044425){7}{\line(1,0){.178074}}
\multiput(152.585,42.625)(.209602,-.043756){6}{\line(1,0){.209602}}
\multiput(153.843,42.362)(.253365,-.042741){5}{\line(1,0){.253365}}
\multiput(155.11,42.149)(.31854,-.04114){4}{\line(1,0){.31854}}
\put(156.384,41.984){\line(1,0){1.2796}}
\put(157.663,41.869){\line(1,0){1.283}}
\put(158.946,41.803){\line(1,0){1.2846}}
\put(160.231,41.787){\line(1,0){1.2843}}
\put(161.515,41.821){\line(1,0){1.282}}
\multiput(162.797,41.905)(.42594,.04431){3}{\line(1,0){.42594}}
\multiput(164.075,42.037)(.31793,.04556){4}{\line(1,0){.31793}}
\multiput(165.347,42.22)(.252746,.04626){5}{\line(1,0){.252746}}
\multiput(166.611,42.451)(.208973,.046667){6}{\line(1,0){.208973}}
\multiput(167.864,42.731)(.177439,.046897){7}{\line(1,0){.177439}}
\multiput(169.106,43.059)(.153556,.047009){8}{\line(1,0){.153556}}
\multiput(170.335,43.435)(.134776,.047033){9}{\line(1,0){.134776}}
\multiput(171.548,43.859)(.119571,.046989){10}{\line(1,0){.119571}}
\multiput(172.744,44.329)(.106967,.046889){11}{\line(1,0){.106967}}
\multiput(173.92,44.844)(.105074,.050991){11}{\line(1,0){.105074}}
\multiput(175.076,45.405)(.094438,.050432){12}{\line(1,0){.094438}}
\multiput(176.209,46.01)(.0853079,.0498889){13}{\line(1,0){.0853079}}
\multiput(177.318,46.659)(.0773636,.0493543){14}{\line(1,0){.0773636}}
\multiput(178.401,47.35)(.0703706,.048822){15}{\line(1,0){.0703706}}
\multiput(179.457,48.082)(.064153,.0482877){16}{\line(1,0){.064153}}
\multiput(180.483,48.855)(.0622375,.0507326){16}{\line(1,0){.0622375}}
\multiput(181.479,49.667)(.0566861,.049978){17}{\line(1,0){.0566861}}
\multiput(182.443,50.516)(.0516713,.0492366){18}{\line(1,0){.0516713}}
\multiput(183.373,51.402)(.0497285,.0511981){18}{\line(0,1){.0511981}}
\multiput(184.268,52.324)(.0505179,.0562055){17}{\line(0,1){.0562055}}
\multiput(185.127,53.28)(.0483064,.0581171){17}{\line(0,1){.0581171}}
\multiput(185.948,54.268)(.0488991,.0636882){16}{\line(0,1){.0636882}}
\multiput(186.73,55.287)(.0494928,.0699004){15}{\line(0,1){.0699004}}
\multiput(187.473,56.335)(.0500919,.0768881){14}{\line(0,1){.0768881}}
\multiput(188.174,57.411)(.0507025,.0848268){13}{\line(0,1){.0848268}}
\multiput(188.833,58.514)(.0473841,.0867242){13}{\line(0,1){.0867242}}
\multiput(189.449,59.642)(.047661,.095866){12}{\line(0,1){.095866}}
\multiput(190.021,60.792)(.04791,.106514){11}{\line(0,1){.106514}}
\multiput(190.548,61.964)(.048131,.119116){10}{\line(0,1){.119116}}
\multiput(191.029,63.155)(.04832,.13432){9}{\line(0,1){.13432}}
\multiput(191.464,64.364)(.048475,.153099){8}{\line(0,1){.153099}}
\multiput(191.852,65.589)(.048592,.176982){7}{\line(0,1){.176982}}
\multiput(192.192,66.827)(.048663,.208517){6}{\line(0,1){.208517}}
\multiput(192.484,68.078)(.048676,.252292){5}{\line(0,1){.252292}}
\multiput(192.728,69.34)(.0486,.31748){4}{\line(0,1){.31748}}
\multiput(192.922,70.61)(.04838,.4255){3}{\line(0,1){.4255}}
\put(193.067,71.886){\line(0,1){1.2812}}
\put(193.163,73.168){\line(0,1){1.8325}}
\put(92.3,167){\vector(-3,4){.5}}
\multiput(130.75,91)(-.062656641604,.03373015873){1594}{\line(-1,0){.062656641604}}
\put(30.3,145.05){\vector(-3,2){.5}}
\multiput(140.5,101.75)(-.033724340176,.045698924731){1446}{\line(0,1){.045698924731}}
\put(160.7,170){\vector(0,1){.5}}
\multiput(160.5,108.25)(.015,2.26666667){27}{\line(0,1){2.26666667}}
\qbezier(30.25,119.5)(78.25,161.125)(196.25,152.25)
\put(159.893,74.893){\line(0,1){.9779}}
\put(159.923,76.849){\line(0,1){.9779}}
\put(159.952,78.805){\line(0,1){.9779}}
\put(159.982,80.761){\line(0,1){.9779}}
\put(160.011,82.717){\line(0,1){.9779}}
\put(160.041,84.673){\line(0,1){.9779}}
\put(160.07,86.629){\line(0,1){.9779}}
\put(160.099,88.585){\line(0,1){.9779}}
\put(160.129,90.541){\line(0,1){.9779}}
\put(160.158,92.496){\line(0,1){.9779}}
\put(160.188,94.452){\line(0,1){.9779}}
\put(160.217,96.408){\line(0,1){.9779}}
\put(160.246,98.364){\line(0,1){.9779}}
\put(160.276,100.32){\line(0,1){.9779}}
\put(160.305,102.276){\line(0,1){.9779}}
\put(160.335,104.232){\line(0,1){.9779}}
\put(160.364,106.188){\line(0,1){.9779}}
\multiput(159.893,74.893)(-.047794,.065564){12}{\line(0,1){.065564}}
\multiput(158.746,76.467)(-.047794,.065564){12}{\line(0,1){.065564}}
\multiput(157.599,78.041)(-.047794,.065564){12}{\line(0,1){.065564}}
\multiput(156.452,79.614)(-.047794,.065564){12}{\line(0,1){.065564}}
\multiput(155.305,81.188)(-.047794,.065564){12}{\line(0,1){.065564}}
\multiput(154.158,82.761)(-.047794,.065564){12}{\line(0,1){.065564}}
\multiput(153.011,84.335)(-.047794,.065564){12}{\line(0,1){.065564}}
\multiput(151.864,85.908)(-.047794,.065564){12}{\line(0,1){.065564}}
\multiput(150.717,87.482)(-.047794,.065564){12}{\line(0,1){.065564}}
\multiput(149.57,89.055)(-.047794,.065564){12}{\line(0,1){.065564}}
\multiput(148.423,90.629)(-.047794,.065564){12}{\line(0,1){.065564}}
\multiput(147.276,92.202)(-.047794,.065564){12}{\line(0,1){.065564}}
\multiput(146.129,93.776)(-.047794,.065564){12}{\line(0,1){.065564}}
\multiput(144.982,95.349)(-.047794,.065564){12}{\line(0,1){.065564}}
\multiput(143.835,96.923)(-.047794,.065564){12}{\line(0,1){.065564}}
\multiput(142.688,98.496)(-.047794,.065564){12}{\line(0,1){.065564}}
\multiput(141.541,100.07)(-.047794,.065564){12}{\line(0,1){.065564}}
\multiput(159.893,74.893)(-.092857,.050794){9}{\line(-1,0){.092857}}
\multiput(158.222,75.808)(-.092857,.050794){9}{\line(-1,0){.092857}}
\multiput(156.551,76.722)(-.092857,.050794){9}{\line(-1,0){.092857}}
\multiput(154.879,77.636)(-.092857,.050794){9}{\line(-1,0){.092857}}
\multiput(153.208,78.551)(-.092857,.050794){9}{\line(-1,0){.092857}}
\multiput(151.536,79.465)(-.092857,.050794){9}{\line(-1,0){.092857}}
\multiput(149.865,80.379)(-.092857,.050794){9}{\line(-1,0){.092857}}
\multiput(148.193,81.293)(-.092857,.050794){9}{\line(-1,0){.092857}}
\multiput(146.522,82.208)(-.092857,.050794){9}{\line(-1,0){.092857}}
\multiput(144.851,83.122)(-.092857,.050794){9}{\line(-1,0){.092857}}
\multiput(143.179,84.036)(-.092857,.050794){9}{\line(-1,0){.092857}}
\multiput(141.508,84.951)(-.092857,.050794){9}{\line(-1,0){.092857}}
\multiput(139.836,85.865)(-.092857,.050794){9}{\line(-1,0){.092857}}
\multiput(138.165,86.779)(-.092857,.050794){9}{\line(-1,0){.092857}}
\multiput(136.493,87.693)(-.092857,.050794){9}{\line(-1,0){.092857}}
\multiput(134.822,88.608)(-.092857,.050794){9}{\line(-1,0){.092857}}
\multiput(133.151,89.522)(-.092857,.050794){9}{\line(-1,0){.092857}}
\multiput(131.479,90.436)(-.092857,.050794){9}{\line(-1,0){.092857}}
\put(110.75,137.5){\vector(-3,4){.107}}\multiput(131.393,108.643)(-.576389,.798611){37}{{\rule{.4pt}{.4pt}}}
\put(135,139){\makebox(0,0)[cc]{$\cU_{I_s^{\rm right}}$}}
\put(96.25,130.75){\makebox(0,0)[cc]{$\cU_{I_s^{\rm left}}$}}
\put(142,124){\makebox(0,0)[cc]{$f(\xii{I_s^{\rm right}})$}}
\put(115.5,112.5){\makebox(0,0)[cc]{$f(\xii{I_s^{\rm left}})$}}
\put(169.75,74.75){\makebox(0,0)[cc]{$\htp^{m,n}$}}
\put(103.75,163.75){\makebox(0,0)[cc]{$C^s_{\vec I_s}$}}
\multiput(130.75,91)(-.062656641604,.03373015873){1594}{\line(-1,0){.062656641604}}
\multiput(140.5,101.75)(-.033724340176,.045698924731){1446}{\line(0,1){.045698924731}}
\multiput(160.5,108.25)(.015,2.26666667){27}{\line(0,1){2.26666667}}
\put(151,136){\circle*{1.736}}
\put(196.75,137){\vector(1,0){.107}}\qbezier(151,136)(174.625,138.5)(194.75,137)
\put(140.5,105.25){\vector(3,-4){.107}}\multiput(110.143,147.393)(.570755,-.79717){54}{{\rule{.4pt}{.4pt}}}
\put(157.5,112){\vector(3,1){.107}}\multiput(140.393,105.143)(.89474,.35526){20}{{\rule{.4pt}{.4pt}}}
\put(157.75,135){\vector(0,1){.107}}\multiput(157.393,111.893)(.01042,.95833){25}{{\rule{.4pt}{.4pt}}}
\put(147,135.55){\vector(-1,0){.107}}\multiput(157.643,134.893)(-.975,.075){12}{{\rule{.4pt}{.4pt}}}
\put(157.75,138.25){\vector(3,1){.107}}\multiput(146.893,136.343)(.88636,.15273){12}{{\rule{.4pt}{.4pt}}}
\put(158,151.25){\vector(0,1){.107}}\multiput(157.643,138.143)(.01786,.92857){15}{{\rule{.4pt}{.4pt}}}
\put(109.75,148.25){\vector(-1,0){.107}}\multiput(157.893,151.143)(-.984694,-.061224){50}{{\rule{.4pt}{.4pt}}}
\end{picture}
\caption{Example of a region in $\CP$ with one closed and few open contours.
Dotted arrows indicate the contours along which one integrates functions of Darboux coordinates resulting from a modular transformation.}
\label{contours}
\end{figure}
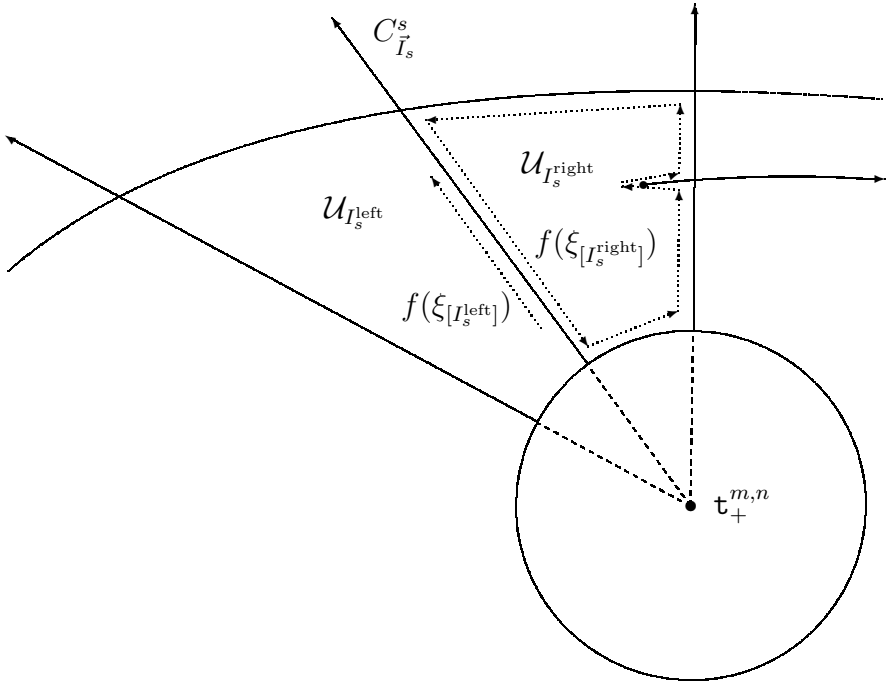

Indeed, let us consider instead of the covering \eqref{gencover} the following set of contours
\be
C_+,\ C_-,\ \{ C_{\vec I_s}^s\},
\qquad
C_{\vec I_s}^s\ \mathop{\mapsto}\limits^{\gl{} }\ C_{\gl{}^{\rm tr} \vec I_s}^s,
\label{gencontour}
\ee
where some of the contours $C_{\vec I_s}^s$ may be open, but one of the invariant sets is still given
by closed contours surrounding the zeros of $c\xi^0+d$. The associated transition functions
are given by $F(\xi)$ and $\bF(\xi)$ for $C_\pm$ and $G_{\vec I_s}^s(\xi,\txi,\alpha)$ for  $C_{\vec I_s}^s$.
Then we claim that all equations proven in this work are still valid for these twistor data.
To understand why this is the case, note that
the set of all contours splits $\CP$ into chambers which replace the usual patches.
We will denote by $\cU_{I_s^{\rm left}}$ and $\cU_{I_s^{\rm right}}$ the neighboring chambers, separated by
the contour $C_{\vec I_s}^s$, on the left and right from it, respectively (see Fig. \ref{contours}).
The next important fact is that all integrals which are evaluated by residues in this work,
and thus the only ones which may be problematic,
appear as a result of an $SL(2,\IZ)$ transformation
and for the twistor data \eqref{gencontour} take the following form
\be
\sum_{s,\vec I_s} \int_{C_{\vec I_s}^s} \frac{\de t'}{2\pi\I t'}
\[f (t',\xii{I_s^{\rm left}} )-f (t',\xii{I_s^{\rm right}} )\].
\label{contrdiff}
\ee
A typical example is given in \eqref{transtxi} where
$f (t',\xi )=-\frac{c}{2}\, K(t,t')\,\frac{\kappa_{abc}\xi^b\xi^c}{c\xi^0+d}$.
Trading the minus sign of the second term for the inversion of the integration contour, one finds that
such contribution is equal to\footnote{One could worry about the situation when some open contours end
at the zeros of $c\xi^0+d$, as shown in Fig. \ref{contours} and indeed happens in physically interesting cases
of D3 and fivebrane instantons \cite{Alexandrov:2012au,Alexandrov:2010ca}, because the formula \eqref{closedint}
seems to ignore the contributions of the integrals along the dashed parts of the contours inside the circle.
This can be justified as follows. The transition function associated with the closed contour
surrounding a zero $\htp^{m,n}$ must ensure the regularity of Darboux coordinates near this point, and therefore
it must cancel the discontinuities generated by the open contours.
This implies that it should have the same branch cut discontinuities along the dashed parts of the contours
as given by the transition functions associated with them.
Taking this into account, one can check that the contributions of the dashed parts cancel in all expressions.
}
\be
\sum_{i} \oint_{C_{i}} \frac{\de t'}{2\pi\I t'}\,f (t',\xii{i} ).
\label{closedint}
\ee
where the sum runs over all chambers and $C_i$ are closed contours surrounding them.
As a result, the integrals can again be evaluated by residues reproducing our previous results.

\subsubsection{Relaxed reality conditions}

Finally, one can relax the reality conditions \eqref{tauU} and \eqref{realG}.
Once the twistor data contain several invariant sets as in \eqref{gencover} or \eqref{gencontour},
it is not necessary anymore to require that each set is preserved by the antipodal map.
It is sufficient if the sets, together with transition functions, are mapped to each other.
Note, however, that the reality condition satisfied by $\xi^0$ ensures that the set comprising the contours surrounding
the zeros of $c\xi^0+d$ is preserved by the action of $\varsigma$.

Thus, the most general twistorial data consistent with the isometric action of the modular group are provided by
the set of contours and transition functions satisfying
\be
\begin{split}
\varsigma\[C_{\vec I_s}^s\]&= C_{-\vec I_{\varsigma(s)}}^{\varsigma(s)},
\qquad
\gl{c,d}\[C_{\vec I_s}^s\]= C_{\gl{}^{\rm tr} \vec I_s}^s,
\\
\overline{\varsigma\[G_{\vec I_s}^s\]}&=G_{-\vec I_{\varsigma(s)}}^{\varsigma(s)},
\qquad
\gl{c,d}\[G_{\vec I_s}^s\]
=\frac{G^s_{(\gl{}^{\rm tr})^{-1}\vec I_s}}{c\xii{((\gl{}^{\rm tr})^{-1}\vec I_s)^{\rm left}}^0+d}
\ +\ {\mbox{non-linear}\atop\mbox{terms}}\ +\ {\mbox{regular}\atop \mbox{terms}}\, ,
\end{split}
\label{StransG}
\ee
where the non-linear terms are as in \eqref{master} with the patch indices $[m',n']$ and $[0]$
replaced by $((\gl{}^{\rm tr})^{-1}\vec I_s)^{\rm left}$ and $((\gl{}^{\rm tr})^{-1}\vec I_s)^{\rm right}$,
respectively, and the regular terms are allowed only for the transition functions
associated to closed contours.

\section{Conclusions}
\label{sec-concl}

In this work, using the twistor approach, we described quaternion-K\"ahler manifolds with the isometry group $SL(2,\IZ)$.
Motivated by a string theory incarnation of this problem,
we restricted ourselves to the manifolds obtained by a deformation of a space
given by the c-map of the moduli space $\KK$ of complexified K\"ahler structures of a Calabi-Yau threefold.
Among our findings, there are three results which we would like to discuss here.

First, we provided a condition on the modular transformation of the transition functions generating deformations
ensuring the presence of the isometric action of $SL(2,\IZ)$. Given in \eqref{master}, the resulting transformation consists
of a simple linear term and very complicated non-linear terms.
The former suggests that the transition functions are similar to coefficients of the Poincar\'e series representation
of a modular form of weight -1, which thus can be probably used to classify solutions to the constraint \eqref{master}.
However, the non-linear terms spoil this simple interpretation and
make it difficult to apply the modular constraint in practice, as was done, for instance,
in \cite{Alexandrov:2010ca} where S-duality was used to derive NS5-brane corrections from D5-instantons in the linear approximation
by the method of images. In principle, the constraint \eqref{master} is exactly the one which should allow to extend this construction
to a non-linear order.

In fact, at least partially, the non-linear terms in \eqref{master} appear due to that transition functions depend on
Darboux coordinates in different patches ($\xi^\Lambda$ are taken in one patch, and $\txi_\Lambda,\alpha$ are from a different one).
This fact significantly complicates the implementation of various symmetries.
For instance, in this formalism the transition functions generating
D-instanton corrections to the HM moduli space are not manifestly symplectic invariant \cite{Alexandrov:2009zh},
despite this symmetry is explicitly realized on Darboux coordinates. In that case it is possible to give an alternative formulation
with transition functions dependent of coordinates from one patch only, which introduces a lot of simplifications.
This possibility is a special feature of the contactomorphisms describing D-instanton corrections to the HM geometry.
One can hope that a similar simplification is also possible for fivebrane instantons and it would remove
(at least some of) the non-linear terms in \eqref{master}.

Our second important result is the change of coordinates \eqref{defYL} and \eqref{qmirror}. As we explained,
being applied to the case where $\cM$ is the HM moduli space,
it is essentially equivalent to the non-perturbative mirror map between type IIA and type IIB physical fields,
generalizing the classical mirror map \eqref{symptobd}.
It is interesting that it is almost trivial for the coordinates appearing as zero modes, which physically correspond
to the RR-fields and the NS5-axion, whereas it is quite non-trivial for the coordinates corresponding
to the CY moduli encoded in $Y^\Lambda$ \eqref{defYL}. It would be interesting to deeper understand
a geometric or physical origin of this map.

Finally, as a byproduct of our analysis, we constructed a modular invariant function on the QK manifold
given explicitly in \eqref{constpartcpchk}. Even the existence of such function,
which encodes all deformations in a non-trivial way, was {\it a priori} non-evident.
In particular, in \cite{Pioline:2009qt,Bao:2009fg} an attempt was done to construct a similar invariant object
(with an extended symmetry group), but its geometric meaning was not clear. Our work suggests that
it should be associated with our function $\Fi$ which has its origin in (but does not coincide with)
the constant part of the contact potential. From physics point of view,
one can also expect that $\Fi$ should have an interpretation as some S-duality invariant partition function.

\section*{Acknowledgments}

We are grateful to Carlos Contou-Carrere, Daniel Persson and Boris Pioline for useful discussions.

\appendix

\section{SL(2,$\IZ)$ transformations}
\label{ap_transform}

\subsection{Useful relations}
\label{subap-rel}

Using \eqref{transftt} and \eqref{invpoints}, one can establish the following transformation properties
\bea
\frac{\de \varpi}{\varpi}\ &\mapsto&\
\frac{(\htp^{c,d}-\htm^{c,d})\de t}{(t-\htp^{c,d})(t-\htm^{c,d})},
\nn
\\
\frac{\de \varpi'}{\varpi'}\,\frac{\varpi'+\varpi}{\varpi'-\varpi}\ &\mapsto&\
\frac{\de \varpi'}{\varpi'}\,\frac{\varpi'+\varpi}{\varpi'-\varpi}-
\frac{\de \varpi'}{\varpi'}\,\frac{\varpi'^2-\htp^{c,d}\htm^{c,d}}{(\varpi'-\htp^{c,d})(\varpi'-\htm^{c,d})},
\nn
\\
\tinvp-\tinvm\ &\mapsto&\
\cC_-^{c,d}\, \frac{(\tinvp-\tinvm)(\htp^{c,d}-\htm^{c,d})}{(\tinvp-\htm^{c,d})(\tinvm-\htm^{c,d})},
\label{transkerroots}
\\
t^{-1}(t-\tinvp)(t-\tinvm)\ &\mapsto&\
\frac{\cC_-^{c,d} (\htp^{c,d}-\htm^{c,d})^2}{(\tinvp-\htm^{c,d})(\tinvm-\htm^{c,d})}
\, \frac{(t-\tinvp)(t-\tinvm)}{(t-\htp^{c,d})(t-\htm^{c,d})},
\nn\\
\frac{\tinvp\tinvm -t^2}{(t-\tinvp)(t-\tinvm)} \ &\mapsto&\
\frac{(\tinvp+\tinvm)(\htp^{c,d}\htm^{c,d}-t^2)-(\htp^{c,d}+\htm^{c,d})(\tinvp\tinvm-t^2)+2t(\tinvp\tinvm-\htp^{c,d}\htm^{c,d})}
{(\htp^{c,d}-\htm^{c,d})(t-\tinvp)(t-\tinvm)}.
\nn
\eea

\subsection{Derivatives of $G_{m,n}$}
\label{ap_trder}

Applying the modular transformation \eqref{SL2Zxi} to the gluing conditions \eqref{QKgluing} between $\cU_0$ and $\cU_{m,n}$,
one finds that the derivatives of $G_{m,n}$ transform as
\bea
\begin{split}
\begin{pmatrix} \p_{\txii{0}_0} G_{m,n} \\ \p_{\ai{0}} G_{m,n} \end{pmatrix}&\ \mapsto\
\frac{1}{c\xii{m',n'}^0+d}
\begin{pmatrix} a & b \\ c & d  \end{pmatrix}
\begin{pmatrix} \p_{\txii{0}_0} G_{m',n'} \\ \p_{\ai{0}} G_{m',n'} \end{pmatrix}
,
\\
\p_{\txii{0}_a}G_{m,n} &\ \mapsto\ \frac{\p_{\txii{0}_a}G_{m',n'}}{c\xii{m',n'}^0+d}
,
\\
\p_{\xi^a_{[m,n]}}G_{m,n} &\ \mapsto\ \p_{\xii{m',n'}^a}G_{m',n'}
+\frac{c}{2}\,\kappa_{abc}\(\frac{\xii{0}^b\xii{0}^c}{c\xii{0}^0+d}-\frac{\xii{m',n'}^b\xii{m',n'}^c}{c\xii{m',n'}^0+d}\)
,
\\
\p_{\xi^0_{[m,n]}}G_{m,n} &\ \mapsto\ d\,\p_{\xii{m',n'}^0}G_{m',n'}-c\(1-\xii{m',n'}^\Lambda\p_{\xii{m',n'}^\Lambda}\)G_{m',n'}
\\
&+\frac{c^2}{6}\,\kappa_{abc}\(\frac{\xii{0}^a\xii{0}^b\xii{0}^c}{c\xii{0}^0+d}-\frac{\xii{m',n'}^a\xii{m',n'}^b\xii{m',n'}^c}{c\xii{m',n'}^0+d}\)
,
\\
(1-\xi^\Lambda_{[m,n]}&\p_{\xi^\Lambda_{[m,n]}})G_{m,n} \ \mapsto\ a\(1-\xii{m',n'}^\Lambda\p_{\xii{m',n'}^\Lambda}\)G_{m',n'}
- b\,\p_{\xii{m',n'}^0}G_{m',n'}
\\
& \hspace{-1cm}
-\frac{ac}{6}\,\kappa_{abc} \(\frac{\xii{0}^a\xii{0}^b\xii{0}^c}{c\xii{0}^0+d} -\frac{\xii{m',n'}^a\xii{m',n'}^b\xii{m',n'}^c}{c\xii{m',n'}^0+d} \)
-\frac{c}{6}\,\kappa_{abc}  \(\frac{\xii{0}^a\xii{0}^b\xii{0}^c}{(c\xii{0}^0+d)^2}-\frac{\xii{m',n'}^a\xii{m',n'}^b\xii{m',n'}^c}{(c\xii{m',n'}^0+d)^2}\)
,
\end{split}
\label{exacttrans}
\eea
where the patch indices $(m',n')$ are defined in \eqref{mpnp}.
In particular, the first transformation implies
\be
1- \p_{\ai{0} }G_{m,n} \ \mapsto\ \frac{c\xii{0}^0+d}{c\xii{m',n'}^0+d}\(1- \p_{\ai{0} }G_{m',n'}\).
\label{trGder-al}
\ee

Given the form of the gluing conditions and the integral equations \eqref{txiqline}, it is convenient to trade
the derivatives $\p_{\txii{0}_\Lambda}G_{m,n}$ in favour of the combination $T_{m,n}^\Lambda$ defined in \eqref{defT}.
In its terms, the transformations \eqref{exacttrans} can be rewritten as
\bea
\begin{split}
\label{transSL2Tex-fact}
T^0_{m,n} &\ \mapsto\ \frac{T^0_{m',n'}}{(c\xii{0}^0+d)(c\xii{m',n'}^0+d)}
=\frac{1}{c(c\xii{0}^0+d)}-\frac{1}{c(c\xii{m',n'}^0+d)}
,
\\
T^a_{m,n} &\ \mapsto\ \frac{T^a_{m',n'}}{c\xii{m',n'}^0+d}
-\frac{c\, \xii{0}^a\, T^0_{m',n'}}{(c\xii{0}^0+d)(c\xii{m',n'}^0+d)}
=\frac{\xii{m',n'}^a}{c\xii{m',n'}^0+d}
-\frac{ \xii{0}^a}{c\xii{0}^0+d}
.
\end{split}
\eea
Combining all these relations, it is straightforward to derive the modular constraint \eqref{master} on the functions $G_{m,n}$.

\section{Transformation of $\varpi$ and the K\"ahler potential}
\label{ap-trK}

In this appendix we would like to study the consequences of the requirement that the K\"ahler potential
on the twistor space \eqref{Knuflat} varies by a K\"ahler transformation under the $SL(2,\IZ)$ action.
Let us evaluate $K_{\cZ}$ for the twistor data defined in section \ref{subsec-trfun}.
Substituting \eqref{solcontpot} and changing the integration kernel to the invariant one \eqref{invkernel2},
the result can be written in the following form
\be
\label{Kahlpot}
K_{\cZ}\ui{0} = \log\frac{1+\varpi\bar \varpi}{|\varpi|}
+ \Re{\sum_{m,n}}' \oint_{C_{m,n}} \frac{\de \varpi'}{2 \pi \I \varpi'}
\,K(\varpi,\varpi') \log\(1- \p_{\ai{0} }G_{m,n}(\varpi')\)+\hat\phi.
\ee
where we fixed for concreteness the patch and
redefined the zero mode to include the shift of the kernel\footnote{At this point of the derivation, $\kk(\varpi)$
is still an unknown function. It can be found only after one fixes the $SL(2,\IZ)$ action on the $\CP$ coordinate $t$.}
\be
\hat\phi=\phi-\hf\,\Re{\sum_{m,n}}' \oint_{C_{m,n}} \frac{\de \varpi'}{2 \pi \I \varpi'}
\,\kk(\varpi') \log\(1- \p_{\ai{0} }G_{m,n}(\varpi')\).
\label{defhatphi}
\ee
Applying an $SL(2,\IZ)$ transformation and using \eqref{trGder-al}, one finds
\be
\begin{split}
\gl{c,d}\[K_{\cZ}\ui{0} \]&= \log\frac{1+\glt{c,d}\overline{\glt{c,d}}}{\left|\glt{c,d}\right|}
+\gl{c,d}[\hat\phi]
\\
&\,
+\Re{\sum_{m,n}}' \oint_{C_{m,n}} \frac{\de \varpi'}{2 \pi \I \varpi'}
\,K(\varpi,\varpi') \[\log\(1- \p_{\ai{0} }G_{m,n}(\varpi')\)+\log\frac{c\xii{0}^0(t')+d}{c\xii{m,n}^0(t')+d}\].
\end{split}
\ee
From this result one concludes that $K_{\cZ}$ transforms by the K\"ahler transformation \eqref{transKZ}
only if the following equation is satisfied
\be
\log\frac{1+\glt{c,d}\overline{\glt{c,d}}}{\left|\glt{c,d}\right|}=\log\frac{1+\varpi\bar \varpi}{|\varpi|}-\Re\Psi_{c,d}(t),
\label{eqKZ-tr}
\ee
where
\be
\Psi_{c,d}(t)=\log\(c\xii{0}^0(t)+d\)+{\sum_{m,n}}' \oint_{C_{m,n}} \frac{\de \varpi'}{2 \pi \I \varpi'}
\,K(\varpi,\varpi')\log\frac{c\xii{0}^0(t')+d}{c\xii{m,n}^0(t')+d}+\gl{c,d}[\hat\phi]-\hat\phi.
\label{defPsi}
\ee
Differentiating \eqref{eqKZ-tr} with respect to $\varpi$ and $\bar\varpi$ and using holomorphicity of $\Psi_{c,d}(t)$,
one obtains the following differential equation
\be
\frac{\left|\p_t\glt{c,d}\right|}{1+\left|\glt{c,d}\right|^2} = \frac{1}{1+|\varpi|^2}.
\ee
Using the original equation \eqref{eqKZ-tr}, it takes a very simple form
\be
\left|\p_t\log \glt{c,d}\right|=\frac{e^{-\Re\Psi_{c,d}(t)}}{|t|},
\label{dereqSt}
\ee
which allows a holomorphic factorization and can be easily integrated. The result reads
\be
\glt{c,d} = \cC\exp \[\int^\varpi \frac{\de{\varpi}}{t} \,
e^{\I\delta-\Psi_{c,d}(t)}\],
\label{resSt}
\ee
where complex $\cC$ and real $\delta$ are independent of $t$.

This integral formula for the transformation of the $\CP$ coordinate can actually be made much more explicit.
This can be achieved by explicitly evaluating the $t$-dependent part of the function $\Psi(t)$ \eqref{defPsi}.
Let us introduce convenient notations
\be
\rhocd(t)=-\frac{t(c\xii{0}^0(t)+d)}{(t-\htp^{c,d})(t-\htm^{c,d})},
\qquad
\rhoi{c,d}(t)=-\frac{t(c\xii{c,d}^0(t)+d)}{(t-\htp^{c,d})(t-\htm^{c,d})}.
\label{defrho}
\ee
According to our assumptions, $\rhoi{c,d}(t)$ is regular and non-vanishing in $\cU_{c,d}$
and the same is true for $\rhocd(t)$ in $\cU_0$. This allows to obtain
\bea
&& \hf\,{\sum\limits_{m,n}}'
\oint_{C_{m,n}}\frac{\de\varpi'}{2\pi \I \varpi'}\, \frac{\varpi'+\varpi}{\varpi'-\varpi}\,
\log\frac{c\xi^0_{[m,n]}+d}{c\xi^0_{[0]}+d}
\nn\\
&=&
\hf\sum_\pm\oint_{C_{\pm c,\pm d}}\frac{\de\varpi'}{2\pi \I \varpi'}\,
\frac{\varpi'+\varpi}{\varpi'-\varpi}\,\log\(\pm\rhoi{\pm c,\pm d}(t)\)
+\hf\,\oint_{C_{0}\cup C_{+}\cup C_{-}}\frac{\de\varpi'}{2\pi \I \varpi'}\,
\frac{\varpi'+\varpi}{\varpi'-\varpi}\,\log\rhocd(\varpi')
\nn
\\
&=& \log\(\rhocd(\varpi)\,\frac{\sqrt{-\htp^{c,d}\htm^{c,d}}}{|c\, Y^0|}\).
\label{evalint}
\eea
Here at the first step we used the regularity of $\log(c\xii{m,n}^0+d)$ in $\cU_{m,n}$ for $(m,n)\ne(c,d)$, and
at the second step we evaluated the contour integrals by residues.
As a result, one gets
\be
\Psi_{c,d}(t)=\log\frac{(t-\htp^{c,d})(t-\htm^{c,d})}{t(\htp^{c,d}-\htm^{c,d})}-\log\gamma_{c,d},
\ee
where
\be
\gamma_{c,d}=\frac{\sqrt{-\htp^{c,d}\htm^{c,d}}}{|c\, Y^0|(\htm^{c,d}-\htp^{c,d})} \, \exp\[{\hat\phi-\gl{c,d}[\hat\phi]
+\hf{\sum_{m,n}}' \oint_{C_{m,n}} \frac{\de \varpi}{2 \pi \I \varpi}
\,\kk(\varpi)\log\frac{c\xii{m,n}^0+d}{c\xii{0}^0+d}}\].
\ee
Substituting this into \eqref{resSt}, one arrives at
\be
\glt{c,d} =\cC\(\frac{t-\htp^{c,d}}{t-\htm^{c,d}}\)^{\gamma_{c,d}+\I\delta}
\ee
Furthermore, one can check that the equation \eqref{eqKZ-tr} is satisfied iff $\gamma_{c,d}+\I\delta=1$
and $|\cC|=|\htm^{c,d}|$.
Given the reality of $\gamma_{c,d}$ which follows from \eqref{realtpm}, this requires $\delta=0$ and $\gamma_{c,d}=1$,
which can be equivalently written as a constraint on the transformation of the zero mode $\hat\phi$,
\be
\gl{c,d}[\hat\phi]=\hat\phi
+\hf{\sum_{m,n}}' \oint_{C_{m,n}} \frac{\de \varpi}{2 \pi \I \varpi}
\,\kk(\varpi)\log\frac{c\xii{m,n}^0+d}{c\xii{0}^0+d}
+\log\frac{\sqrt{-\htp^{c,d}\htm^{c,d}}}{|c\, Y^0|(\htm^{c,d}-\htp^{c,d})}.
\label{transf-hatphi}
\ee
The resulting formula for $\glt{c,d}$ coincides with \eqref{transftt} where we slightly changed the notation
for the prefactor.

\section{Proof of modular properties}
\label{ap_check}

\subsection{Proof of the Lemma}
\label{ap_Lemma}

The proof of the Lemma presented in section \ref{subsec-mirmap} does not require any non-trivial steps and can be achieved
by consistently applying $SL(2,\IZ)$ transformations to all ingredients entering the definition of $Y^\Lambda$ \eqref{defYL}.
Therefore, we present here just the main steps of this calculation.

First, note that the transformation of the kernel appearing in \eqref{defYL}
\be
M(t)\equiv\gl{c,d}\[\frac{\tinvp\tinvm}{2}\,\frac{\de \varpi}{2 \pi \I}\,
\frac{ \tinvp(\tinvm-t)+\tinvm(\tinvp-t)}{(t-\tinvp)^2(t-\tinvm)^2}\]
\ee
has poles only at critical point $\tinvpm$ with residues given by
\bea
& \mathop{\rm Res}\limits_{t=\tinvpm}\[M(t)f(t)\]=
\pm\frac{\Cfm}{2}\, \frac{(\tinvp-\htp^{c,d})(\tinvm-\htp^{c,d})}{(\tinvp-\tinvm)(\htp^{c,d}-\htm^{c,d})}
\(f(\tinvpm)-\frac{(\tinvpm-\htp^{c,d})(\tinvpm-\htm^{c,d})}{\htp^{c,d}-\htm^{c,d}}\,\p_t f(\tinvpm)\), &
\nn\\
& \mathop{\rm Res}\limits_{t=\htp^{c,d}}\[\frac{M(t)f(t)}{t-\htp^{c,d}}\]=
\frac{\Cfm f(\htp^{c,d})}{\htp^{c,d}-\htm^{c,d}},
\qquad\qquad
\mathop{\rm Res}\limits_{t=\htm^{c,d}}\[\frac{M(t)f(t)}{t-\htm^{c,d}}\]= 0. &
\label{residues}
\eea
for any regular function $f(t)$. These properties are already sufficient to reproduce the transformation of $Y^0$ \eqref{transfYL}.
Indeed, using \eqref{transSL2Tex-fact}, one finds that the transformed integral in \eqref{defYL} can be evaluated by taking residues
at $\tinvpm$ and $\htpm^{c,d}$, which all can be computed using \eqref{residues}.
Then, taking into account the conditions \eqref{cond-spm} and the characteristic equation \eqref{eqinvpoints} for the critical points,
the former can be checked to cancel the contribution of the first term in \eqref{defYL}, whereas the latter provide the desired result.

On the other hand, for $Y^a$ one obtains
\bea
\gl{c,d}[Y^a]&=&- \frac{\Cfm\htp^{c,d}\xii{c,d}^a(\htp^{c,d})}{(\htp^{c,d} - \htm^{c,d})^2 \rhoi{c,d}(\htp^{c,d})}
-\frac{\Cfm}{2}\, \frac{(\tinvp-\htp^{c,d})(\tinvm-\htp^{c,d})}{(\tinvp-\tinvm)(\htp^{c,d}-\htm^{c,d})}
\Biggr[\frac{2\I\tau_2}{|c\tau+d|^2}\(\gl{c,d}[\cY^a]-(cc^a+db^a)\)
\Biggr.
\nn\\
&&\Biggl.
-\frac{1}{\htp^{c,d}-\htm^{c,d}}\(\frac{\tinvp\p_t\xii{0}^a(\tinvp)}{\rhocd(\tinvp)}-\frac{\tinvm\p_t\xii{0}^a(\tinv,)}{\rhocd(\tinvm)}\)\Biggr],
\label{inter-resYa}
\eea
where here and below we use extensively the notations introduced in \eqref{defrho}.
To get $\gl{c,d}[\cY^a]$, one finds
\bea
&&\gl{c,d}[U]
=U+\frac{\I}{2}\,\log\frac{\rhocd(\tinvp)}{\rhocd(\tinvm)},
\\
&&
\begin{split}
\gl{c,d}[\cI^a]=&\,
\frac{\tau_2(cc^a+db^a)}{|c\tau+d|^2}\,
\frac{2(\tinvp\tinvm+\htp^{c,d}\htm^{c,d})-(\tinvp+\tinvm)(\htp^{c,d}+\htm^{c,d})}{(\tinvp-\tinvm)(\htp^{c,d}-\htm^{c,d})}
\\
&\,
-\frac{\I}{2(\htp^{c,d}-\htm^{c,d})}
\(\frac{\tinvp\p_t\xii{0}^a(\tinvp)}{\rhocd(\tinvp)}+\frac{\tinvm\p_t\xii{0}^a(\tinv,)}{\rhocd(\tinvm)}\),
\end{split}
\eea
where the transformed integrals are evaluated using the same strategy as in \eqref{evalint}
with help of the conditions \eqref{cond-spm} and \eqref{eqinvpoints}.
Applying these results, one can show that the second term in \eqref{inter-resYa} is proportional to
\be
e^{-\I U}\tinvp\p_t\xii{0}^a(\tinvp)-e^{\I U}\tinvm\p_t\xii{0}^a(\tinvm)+4\I\tau_2 t^a\, \frac{ \sqrt{\tinvp\tinvm}}{\tinvp-\tinvm}
\ee
which vanishes due to the condition \eqref{cond-spmder}. As a result, one remains with the first term only in \eqref{inter-resYa}
which precisely reproduces \eqref{transfYL}.

\subsection{Proof of the Theorem}
\label{ap_Theorem}

To prove the Theorem of section \ref{subsec-mirmap},
let us first write explicitly the integral equations \eqref{txiqline} after substitution of the mirror map \eqref{qmirror}
and the transition functions \eqref{transfun}. Choosing for concreteness $t\in\cU_0$ and
evaluating the integrals corresponding to the patches $\cU_\pm$, one obtains the equations in the following form
\bea
\hspace{-0.5cm}
\xii{0}^\Lambda(t) &=& \zeta^\Lambda_{\rm cl}
-\hf\, (\tinvp+\tinvm)\(\frac{Y^\Lambda}{\tinvp\tinvm}-\bY^\Lambda\)+\frac{Y^\Lambda}{t}-t\bY^\Lambda
+ {\sum_{m,n}}' \oint_{C_{m,n}} \frac{\de \varpi'}{2 \pi \I \varpi'} \,
K(\varpi,\varpi')\, T_{m,n}^\Lambda,
\nn
\\
\hspace{-0.5cm}
\txii{0}_\Lambda(t) &=& \tzeta_\Lambda^{\rm cl}
-\hf\, (\tinvp+\tinvm)\(\frac{F_\Lambda(Y)}{\tinvp\tinvm}-\bF_\Lambda(\bY)\)+\frac{F_\Lambda(Y)}{t}-t\bF_\Lambda(\bY)
\nn\\
&&
-{\sum_{m,n}}' \oint_{C_{m,n}} \frac{\de \varpi'}{2 \pi \I \varpi'} \,
K(\varpi,\varpi')\, \p_{\xii{m,n}^\Lambda}G_{m,n},
\label{qmirrorG}
\\
\hspace{-0.5cm}
\ai{0}(t) &=& (\sigma+\zeta^\Lambda \tzeta_\Lambda)^{\rm cl}
+\hf\, (\tinvp^2+\tinvm^2)\(\frac{F(Y)}{(\tinvp\tinvm)^{2}} + \bF(\bY) \)-\(\frac{F(Y)}{\varpi^2} + \varpi^2 \bF(\bY) \)
\nn\\
&&
+\(\frac{\tinvp+\tinvm}{2\tinvp\tinvm}-\frac{1}{\varpi} \) F_\Lambda(Y)\xi_0^\Lambda
-\(\frac{\tinvp+\tinvm}{2}-\varpi \) \bF_\Lambda(\bY)\bar\xi_0^\Lambda
\nn\\
&&
- {\sum_{m,n}}' \oint_{C_{m,n}}\frac{\de \varpi'}{2 \pi \I \varpi'}
K(\varpi,\varpi')\(1-{\xii{m,n}^\Lambda} (\varpi')\partial_{{\xii{m,n}^\Lambda}}\)G_{m,n} ,
\nn
\eea
where
\be
\xi_0^\Lambda=\zeta^\Lambda_{\rm cl}
-\hf\, (\tinvp+\tinvm)\(\frac{Y^\Lambda}{\tinvp\tinvm}-\bY^\Lambda\)
+\hf{\sum_{m,n}}' \oint_{C_{m,n}} \frac{\de \varpi'}{2 \pi \I \varpi'} \,
\frac{\tinvp(\tinvm-\varpi')+\tinvm(\tinvp-\varpi')}{(\varpi'-\tinvp)(\varpi'-\tinvm)}\, T_{m,n}^\Lambda
\ee
is the zeroth term in the Laurent expansion of $\xii{0}^\Lambda$ around $t=0$.

The calculation of the modular transformations of these expressions is greatly simplified
if one uses the Lemma which provides the transformation of $Y^\Lambda$.
To use it, however, one needs to verify the conditions \eqref{cond-spm} and \eqref{cond-spmder}, which in turn requires the explicit form of
the conjugate variables $\bY^\Lambda$. They can be found by taking into account the behaviour
of the critical points and transition functions under the antipodal map (see \eqref{realG}). As a result, one obtains
\be
\bY^\Lambda=
\frac{ \I\tau_2}{\tinvm-\tinvp}\,\bar\cY^\Lambda+
\frac{1}{2}{\sum_{m,n}}' \oint_{C_{m,n}}\frac{\de \varpi}{2 \pi \I}\,
\frac{ t(2t-\tinvp-\tinvm)}{(t-\tinvp)^2(t-\tinvm)^2}\,T^\Lambda_{m,n},
\label{defbYL}
\ee
where
\be
\bar\cY^0=1,
\qquad
\bar\cY^a =
\frac{b^a\(\tinvm e^{-\I U}-\tinvp e^{\I U}\)+2\sqrt{\tinvp \tinvm}\,t^a}{(\tinvm-\tinvp)\cos U}
- \I\tau_2^{-1}\tan U \,\cI^a
\ee
Then a simple calculation gives
\be
\begin{split}
&
\qquad\qquad
\xii{0}^0(\tinvpm)=\zeta^0_{\rm cl} \pm \I\tau_2,
\qquad
\xii{0}^a(\tinvpm)=\zeta^a_{\rm cl} \pm \I\tau_2 b^a,
\\
&
\tinvpm\p_t\xii{0}^a(\tinvpm)= \pm\frac{2\I\tau_2 t^a}{\cos U}\, \frac{ \sqrt{\tinvp\tinvm}}{\tinvm-\tinvp}
-\frac{\I\, e^{\pm \I U}}{\cos U}\(\cI^a-\tau_2 b^a\, \frac{\tinvp+\tinvm}{\tinvp-\tinvm}\),
\end{split}
\ee
which confirmes the required conditions and identifies the vector $L^a$ as
\be
L^a=\frac{1}{\cos U}\(\cI^a-\tau_2 b^a\, \frac{\tinvp+\tinvm}{\tinvp-\tinvm}\)
\ee

It is now straightforward to evaluate the $SL(2,\IZ)$ transformations of the integral equations \eqref{qmirrorG}.
The calculation proceeds in the same way for all equations and therefore we will not repeat it here for all of them.
Let us just describe how it works for the equation involving $\txi_a$
which allows to illustrate the typical manipulations leading to the desired result.

Using \eqref{exacttrans} and the invariance of the kernel, one obtains the transformation of the integral term
\be
{\sum_{m,n}}' \oint_{C_{m,n}} \frac{\de \varpi'}{2 \pi \I \varpi'} \,
K(\varpi,\varpi')\[ \p_{\xii{m,n}^a}G_{m,n}
+\frac{c}{2}\,\kappa_{abc}\(\frac{\xii{0}^b\xii{0}^c}{c\xii{0}^0+d}-\frac{\xii{m,n}^b\xii{m,n}^c}{c\xii{m,n}^0+d}\)\].
\label{transtxi}
\ee
The first term reproduces the integral contribution one started with, whereas the second term can be evaluated by residues.
To this end, the sum over patches $\cU_{m,n}$ for the first term in the round brackets should be rewritten as
an integral over $\cU_0\cup\cU_+\cup\cU_-$ and is equal to
the contributions of the residues at $t'=t$, $t'=\tinvpm$, $t'=0$, and $t'=\infty$.
The first one reproduces the inhomogeneous term in the transformation of $\txi_a$ \eqref{SL2Zxi},
the next two are equal to $\tzeta_a^{\rm cl}-\gl{c,d}[\tzeta_a^{\rm cl}]$, whereas the last two give the terms
with the prepotential in $\txi_a$ \eqref{qmirrorG}. On the other hand, the second term in the round brackets
is given by residues at $\htpm^{c,d}$ and cancels the transformation of these prepotential dependent terms.
As a result, the integral equation turns out be invariant confirming the transformation law of Darboux coordinates.

\subsection{Invariant potential}
\label{ap_zeromode}

Let us represent the potential \eqref{constpartcpchk} as a product of three factors
\be
\Fi=\frac{A\cV}{\cos\Theta},
\ee
where $\cV$ denotes the expression in the curly brackets, $\Theta$ is the integral appearing in the argument of cosine,
and $A$ denotes all the remaining factors. It is straightforward to show the following properties
\bea
\gl{c,d}[\Theta]&=& \frac{\I}{2}\(\Phip-\Phim-\log \frac{\rhoip}{\rhoim}\),
\label{trTheta}\\
\gl{c,d}[A]&=& A\, e^\phi\,(\htm^{c,d}-\htp^{c,d}) \sqrt{\rhoip\rhoim}\,
e^{-\hf \(\Phip+\Phim\)} ,
\label{trA}
\eea
where we denoted $\Phipm=\Phi^{[\pm c,\pm d]}(\htpm^{c,d})$ and $\rhoipm=\pm\rhoi{\pm c,\pm d}(\htpm^{c,d})$.
The $SL(2,\IZ)$ transformation of $\cV$ is more complicated.
It can be significantly simplified by using the following identity
\bea
&&
\sum_j \oint_{C_j}\frac{\de\varpi}{\varpi}\Biggr[\varpi \p_\varpi f \, \Hij{j0}-4\I f(\varpi)  \,e^{\Phi^{[0]}} \p_{\ai{0} }\hHij{j0}
\Biggl.
\label{totder}\\
&&
\left.
-f(\varpi)\p_{\xii{j}^\Lambda}\hHij{j0}\(\varpi^{-1} Y^\Lambda+\varpi\bY^\Lambda-
\sum_k \oint_{C_k}\frac{\varpi\,\de\varpi'}{2\pi\I (\varpi'-\varpi)^2}\,
 \(\p_{\txii{0}_\Lambda }\hHij{k0}
-\xii{0}^\Lambda(\varpi') \, \p_{\ai{0} }\hHij{k0}\)\)\]=0,
\nn
\eea
which holds for any function $f(t)$ just because the l.h.s. is a a total derivative.
In our case, one should choose this function as
\be
f(t)=\frac{1}{32\pi(\htp^{c,d}-\htm^{c,d})}
\(\frac{t+\htp^{c,d} }{(t-\htp^{c,d})\rhoip}+\frac{t+\htm^{c,d}}{(t-\htm^{c,d})\rhoim} \)
\ee
and add \eqref{totder} to the transformation of $\cV$. Then a lengthy calculation yields
\be
\gl{c,d}[\cV]=\frac{\rhoip+\rhoim}{2(\htm^{c,d}-\htp^{c,d})\rhoip\rhoim}\,\cV
-4\I\,{\sum_{m,n}}' \oint_{C_{m,n}} \frac{\de\varpi}{\varpi} f(\varpi)  \,e^{\Phi^{[0]}} \p_{\ai{0} }G_{m,n}.
\ee
The last term can actually be evaluated explicitly. To this end, it can be represented as
\be
\begin{split}
&
4\I\,{\sum_{m,n}}' \oint_{C_{m,n}} \frac{\de\varpi}{\varpi} f(\varpi)  \,e^{\Phi^{[m,n]}}
+ 4\I \oint_{C_0\cup C_{+}\cup C_-} \frac{\de\varpi}{\varpi} f(\varpi)  \,e^{\Phi^{[0]}}
\\ & \qquad\qquad\qquad\qquad\qquad
=\frac{\rhoip\, e^{\Phim}+\rhoim\, e^{\Phip}-(\rhoip+\rhoim) e^\phi \cos\Theta}{2(\htm^{c,d}-\htp^{c,d})\rhoip\rhoim}.
\end{split}
\ee
Taking into account that $ e^\phi \cos\Theta=\cV$, one finds a very simple result
\be
\gl{c,d}[\cV]=\frac{\rhoip\, e^{\Phim}+\rhoim\, e^{\Phip}}{2(\htm^{c,d}-\htp^{c,d})\rhoip\rhoim}.
\ee
Combining it with \eqref{trTheta} and \eqref{trA}, one immediately obtains that the potential $\Fi$ is modular invariant.

\providecommand{\href}[2]{#2}\begingroup\raggedright\endgroup


\begin{thebibliography}{10}

\bibitem{Seiberg:1994rs}
N.~Seiberg and E.~Witten, ``{Monopole Condensation, And Confinement In N=2
  Supersymmetric Yang-Mills Theory},'' {\em Nucl. Phys.} {\bf B426} (1994)
  19--52,
\href{http://www.arXiv.org/abs/hep-th/9407087}{{\tt hep-th/9407087}}.

\bibitem{Seiberg:1996nz}
N.~Seiberg and E.~Witten, ``Gauge dynamics and compactification to three
  dimensions,''
\href{http://www.arXiv.org/abs/hep-th/9607163}{{\tt hep-th/9607163}}.

\bibitem{Bagger:1983tt}
J.~Bagger and E.~Witten, ``{M}atter couplings in {${\mathcal N}=2$}
  supergravity,'' {\em Nucl. Phys.} {\bf B222} (1983)
1.

\bibitem{Green:1997tv}
M.~B. Green and M.~Gutperle, ``{Effects of D-instantons},'' {\em Nucl. Phys.}
  {\bf B498} (1997) 195--227,
\href{http://www.arXiv.org/abs/hep-th/9701093}{{\tt hep-th/9701093}}.

\bibitem{RoblesLlana:2006is}
D.~Robles-Llana, M.~Ro\v{c}ek, F.~Saueressig, U.~Theis, and S.~Vandoren,
  ``{Nonperturbative corrections to 4D string theory effective actions from
  SL(2,Z) duality and supersymmetry},'' {\em Phys. Rev. Lett.} {\bf 98} (2007)
  211602,
\href{http://www.arXiv.org/abs/hep-th/0612027}{{\tt hep-th/0612027}}.

\bibitem{Alexandrov:2011va}
S.~Alexandrov, ``{Twistor Approach to String Compactifications: a Review},''
  {\em Phys.Rept.} {\bf 522} (2013) 1--57,
\href{http://www.arXiv.org/abs/1111.2892}{{\tt 1111.2892}}.

\bibitem{Alexandrov:2013yva}
S.~Alexandrov, J.~Manschot, D.~Persson, and B.~Pioline, ``{Quantum
  hypermultiplet moduli spaces in N=2 string vacua: a review},''
\href{http://www.arXiv.org/abs/1304.0766}{{\tt 1304.0766}}.

\bibitem{Cecotti:1989qn}
S.~Cecotti, S.~Ferrara, and L.~Girardello, ``Geometry of type {I}{I}
  superstrings and the moduli of superconformal field theories,'' {\em Int. J.
  Mod. Phys.} {\bf A4} (1989)
2475.

\bibitem{Ferrara:1989ik}
S.~Ferrara and S.~Sabharwal, ``{Q}uaternionic manifolds for type {II}
  superstring vacua of {C}alabi-{Y}au spaces,'' {\em Nucl. Phys.} {\bf B332}
  (1990)
317.

\bibitem{MR664330}
S.~M. Salamon, ``Quaternionic {K}\"ahler manifolds,'' {\em Invent. Math.} {\bf
  67} (1982), no.~1, 143--171.

\bibitem{Becker:1995kb}
K.~Becker, M.~Becker, and A.~Strominger, ``Five-branes, membranes and
  nonperturbative string theory,'' {\em Nucl. Phys.} {\bf B456} (1995)
  130--152,
\href{http://www.arXiv.org/abs/hep-th/9507158}{{\tt hep-th/9507158}}.

\bibitem{Alexandrov:2012bu}
S.~Alexandrov and B.~Pioline, ``{S-duality in Twistor Space},'' {\em JHEP} {\bf
  1208} (2012) 112,
\href{http://www.arXiv.org/abs/1206.1341}{{\tt 1206.1341}}.

\bibitem{Alexandrov:2008gh}
S.~Alexandrov, B.~Pioline, F.~Saueressig, and S.~Vandoren, ``{D-instantons and
  twistors},'' {\em JHEP} {\bf 03} (2009) 044,
\href{http://www.arXiv.org/abs/0812.4219}{{\tt 0812.4219}}.

\bibitem{Alexandrov:2009zh}
S.~Alexandrov, ``{D-instantons and twistors: some exact results},'' {\em J.
  Phys.} {\bf A42} (2009) 335402,
\href{http://www.arXiv.org/abs/0902.2761}{{\tt 0902.2761}}.

\bibitem{Alexandrov:2010ca}
S.~Alexandrov, D.~Persson, and B.~Pioline, ``{Fivebrane instantons, topological
  wave functions and hypermultiplet moduli spaces},'' {\em JHEP} {\bf 1103}
  (2011) 111, \href{http://www.arXiv.org/abs/1010.5792}{{\tt 1010.5792}}.

\bibitem{Neitzke:2007ke}
A.~Neitzke, B.~Pioline, and S.~Vandoren, ``{Twistors and Black Holes},'' {\em
  JHEP} {\bf 04} (2007) 038,
\href{http://www.arXiv.org/abs/hep-th/0701214}{{\tt hep-th/0701214}}.

\bibitem{Alexandrov:2008nk}
S.~Alexandrov, B.~Pioline, F.~Saueressig, and S.~Vandoren, ``{Linear
  perturbations of quaternionic metrics},'' {\em Commun. Math. Phys.} {\bf 296}
  (2010) 353--403,
\href{http://www.arXiv.org/abs/0810.1675}{{\tt 0810.1675}}.

\bibitem{Robles-Llana:2006ez}
D.~Robles-Llana, F.~Saueressig, and S.~Vandoren, ``String loop corrected
  hypermultiplet moduli spaces,'' {\em JHEP} {\bf 03} (2006) 081,
\href{http://www.arXiv.org/abs/hep-th/0602164}{{\tt hep-th/0602164}}.

\bibitem{Alexandrov:2007ec}
S.~Alexandrov, ``{Quantum covariant c-map},'' {\em JHEP} {\bf 05} (2007) 094,
\href{http://www.arXiv.org/abs/hep-th/0702203}{{\tt hep-th/0702203}}.

\bibitem{Alexandrov:2009qq}
S.~Alexandrov and F.~Saueressig, ``{Quantum mirror symmetry and twistors},''
  {\em JHEP} {\bf 09} (2009) 108,
\href{http://www.arXiv.org/abs/0906.3743}{{\tt 0906.3743}}.

\bibitem{MR872143}
K.~Galicki, ``A generalization of the momentum mapping construction for
  quaternionic {K}\"ahler manifolds,'' {\em Comm. Math. Phys.} {\bf 108}
  (1987), no.~1, 117--138.

\bibitem{Rocek:2005ij}
M.~Ro\v{c}ek, C.~Vafa, and S.~Vandoren, ``Hypermultiplets and topological
  strings,'' {\em JHEP} {\bf 02} (2006) 062,
\href{http://www.arXiv.org/abs/hep-th/0512206}{{\tt hep-th/0512206}}.

\bibitem{Rocek:2006xb}
M.~Ro\v{c}ek, C.~Vafa, and S.~Vandoren, ``{Q}uaternion-{K\"a}hler spaces,
  hyperk{\"a}hler cones, and the c-map,''
\href{http://www.arXiv.org/abs/math.dg/0603048}{{\tt math.dg/0603048}}.

\bibitem{Karlhede:1984vr}
A.~Karlhede, U.~Lindstr{\"o}m, and M.~Ro\v{c}ek, ``Selfinteracting tensor
  multiplets in {$\N=2$} superspace,'' {\em Phys. Lett.} {\bf B147} (1984)
297.

\bibitem{Hitchin:1986ea}
N.~J. Hitchin, A.~Karlhede, U.~Lindstr{\"o}m, and M.~Ro\v{c}ek,
  ``Hyperk{\"a}hler metrics and supersymmetry,'' {\em Commun. Math. Phys.} {\bf
  108} (1987)
535.

\bibitem{Lindstrom:1987ks}
U.~Lindstr{\"o}m and M.~Ro\v{c}ek, ``New hyperkahler metrics and new
  supermultiplets,'' {\em Commun. Math. Phys.} {\bf 115} (1988)
21.

\bibitem{Bohm:1999uk}
R.~B{\"o}hm, H.~G{\"u}nther, C.~Herrmann, and J.~Louis, ``{Compactification of
  type IIB string theory on Calabi-Yau threefolds},'' {\em Nucl. Phys.} {\bf
  B569} (2000) 229--246,
\href{http://www.arXiv.org/abs/hep-th/9908007}{{\tt hep-th/9908007}}.

\bibitem{Alexandrov:2012au}
S.~Alexandrov, J.~Manschot, and B.~Pioline, ``{D3-instantons, Mock Theta Series
  and Twistors},'' {\em JHEP} {\bf 1304} (2013) 002,
\href{http://www.arXiv.org/abs/1207.1109}{{\tt 1207.1109}}.

\bibitem{Alexandrov:2008ds}
S.~Alexandrov, B.~Pioline, F.~Saueressig, and S.~Vandoren, ``{Linear
  perturbations of Hyperkahler metrics},'' {\em Lett. Math. Phys.} {\bf 87}
  (2009) 225--265,
\href{http://www.arXiv.org/abs/0806.4620}{{\tt 0806.4620}}.

\bibitem{ks}
M.~Kontsevich and Y.~Soibelman, ``{Stability structures, motivic
  Donaldson-Thomas invariants and cluster transformations},''
  \href{http://www.arXiv.org/abs/0811.2435}{{\tt 0811.2435}}.

\bibitem{Gaiotto:2008cd}
D.~Gaiotto, G.~W. Moore, and A.~Neitzke, ``{Four-dimensional wall-crossing via
  three-dimensional field theory},'' {\em Commun.Math.Phys.} {\bf 299} (2010)
  163--224, \href{http://www.arXiv.org/abs/0807.4723}{{\tt 0807.4723}}.

\bibitem{Alexandrov:2011ac}
S.~Alexandrov, D.~Persson, and B.~Pioline, ``{Wall-crossing, Rogers
  dilogarithm, and the QK/HK correspondence},'' {\em JHEP} {\bf 1112} (2011)
  027,
\href{http://www.arXiv.org/abs/1110.0466}{{\tt 1110.0466}}.

\bibitem{Pioline:2009qt}
B.~Pioline and D.~Persson, ``{The automorphic NS5-brane},'' {\em Commun. Num.
  Th. Phys.} {\bf 3} (2009), no.~4, 697--754,
\href{http://www.arXiv.org/abs/0902.3274}{{\tt 0902.3274}}.

\bibitem{Bao:2009fg}
L.~Bao, A.~Kleinschmidt, B.~E.~W. Nilsson, D.~Persson, and B.~Pioline,
  ``{Instanton Corrections to the Universal Hypermultiplet and Automorphic
  Forms on SU(2,1)},'' {\em Commun. Num. Theor. Phys.} {\bf 4} (2010) 187--266,
\href{http://www.arXiv.org/abs/0909.4299}{{\tt 0909.4299}}.

\end{thebibliography}

\end{document}